%% file: main.tex
    \documentclass[10pt, final, twocolumn]{extarticle}
    
    \usepackage[]{geometry}
    \usepackage[utf8]{inputenc}
    \usepackage[british]{babel}

    \usepackage{xcolor}
    
    \usepackage{amsmath}
    \usepackage{amssymb}

    \usepackage{bm}
 
    \usepackage[binary-units=true]{siunitx} 

    \allowdisplaybreaks    
    
    \usepackage[font={footnotesize},labelsep=period]{caption}   
    
    \usepackage{pgfplots}\pgfplotsset{compat=1.14}
    \usepackage{PGFplot/PGFplotTUDA/pgfplots-private}

    \usepgfplotslibrary{units} 
    \usepgfplotslibrary{fillbetween}

\input{PGFplot/FIG_Verification.tex}
\input{PGFplot/FIG_Result.tex}
\input{PGFplot/FIG_Other.tex}

    \def\figWidth{4.6cm}  
    \def\figHeight{3.5cm}  
    \def\MyClassDependentFigResultSpan{0.485\columnwidth}

    \usepackage{xr}

    \usepackage{url}
    \usepackage{booktabs}
    
    \usepackage{etoolbox}
    \AtBeginEnvironment{tabular}{\footnotesize}

\input{Table/TAB_SourceCode.tex}

\input{Util/UTIL_mathGreekVariablesV0_0.tex}
\input{Util/UTIL_mathOperatorsV0_0.tex}

\input{Util/UTIL_mathEquations.tex}

\newcommand{\FEATHER}[0]{\mbox{Feather-M2.1-2}}

    \usepackage{authblk}

\newcommand{\MyClassDependentMakeTitleCommand}[0]{

    \title{Numerical Analysis of the Screening Current-Induced Magnetic Field in the HTS Insert Dipole Magnet Feather-M2.1-2}

	\author{\small
			L.~Bortot$^{1,2}$,
			M.~Mentink$^{1}$,
			C.~Petrone$^{1}$,
			J.~Van~Nugteren$^{1}$,
			G.~Kirby$^{1}$,
			M.~Pentella$^{1,3}$,
			A.P.~Verweij$^{1}$,
			and~S.~Sch{\"o}ps$^{2}$
		}
		
    \setlength{\affilsep}{0.5em}

    \affil{$^{1}$ CERN, Espl. des Particules 1, 1211 Geneva, CH}
    \affil{$^{2}$ Technische Universit{\"a}t Darmstadt, Karolinenplatz 5, 64289 Darmstadt, DE}
    \affil{$^{3}$ Department of Applied Science and Technology, Polytechnic of Turin, Turin, IT}    
    \affil{\texttt{{lorenzo.bortot@cern.ch}}}

    \date{}    
    \maketitle
}

\newcommand{\MyClassDependentMakeBibliographyCommand}[0]{
    {\footnotesize  
        \bibliography{Bib/BIB_article,Bib/BIB_book,Bib/BIB_misc,Bib/BIB_thesis}
    }
}

\begin{document}


\MyClassDependentMakeTitleCommand{}


\input{Section/SEC01_Abstract}


\input{Section/SEC02_Body}

\input{Section/SEC03_Acknowledgements}


\MyClassDependentMakeBibliographyCommand{}
 
              
\end{document}

%% file: PGFplot/FIG_Verification.tex
\newcommand{\FigAClossQBparamN}[0]{
    \begin{tikzpicture}
		\begin{loglogaxis}[	
            temflineplot,
            legend pos  = north west,
            width       = 8cm, 
            height      = 6cm,
            xmin        = 1e-4, 
            xmax        = 1e2,
            ymin        = 1e-10, 
            ymax        = 1e+2,	
            ytickten    = {-10,...,+2},
            xlabel      = $\SCAl{B}{\RM{p}}$, 
            ylabel      = $\SCAl{w}{\RM{J}}$,
            x unit      = \si{\tesla}, 
            y unit      = \si{\joule\per\cubic\mm/cycle},
            ]  
            \addplot+ 
                table[x=x,y=n5,col sep=comma] 
                {Data/ACloss_SIM_Q_B_f1Hz_paramN.csv}; 
            \addplot+ 
                table[x=x,y=n20,col sep=comma] 
                {Data/ACloss_SIM_Q_B_f1Hz_paramN.csv}; 
            \addplot+ 
                table[x=x,y=n40,col sep=comma] 
                {Data/ACloss_SIM_Q_B_f1Hz_paramN.csv};
           \addplot+ 
                [mark=none, dashed, thin, gray!100] 
                table[x=xBrandt,y=nBrandt,col sep=comma] 
                {Data/ACloss_SIM_Q_B_f1Hz_paramN.csv};
           \addplot
                [mark=none, dotted, thin, black] 
                coordinates {(1e-4,1e-10) (4e-2,6.4e-3)};
            \node 
                at (4e-4, 5e-7)
                {$\propto\SCAu{B}{3}$}; 
            \node 
                at (4e-3, 2e-9)
                {$\propto\SCAu{B}{4}$};    
            \node 
                at (2e1, 2e0) 
                {$\propto\SCAu{B}{1}$};	
             \addplot+ 
                [mark=none, dotted, thin, black] 
                coordinates {(40e-3,1e-10) (40e-3,1e2)}; 
            \node 
                [black] 
                at (8e-2, 1e1) 
                {$\textrm{B}_\textrm{c}$};			
			\node 
			    [above right] at (rel axis cs:0.52,0.025) 
			    {\fcolorbox{black}{white}{\includegraphics[width=2.5cm]{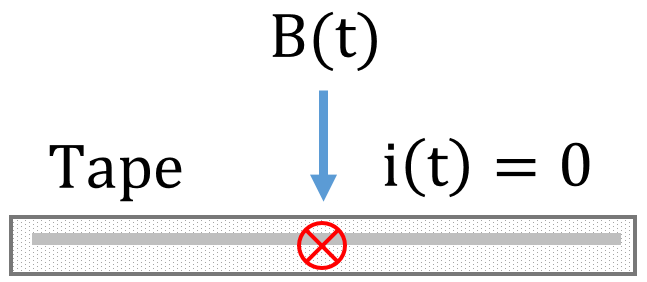}}}; 
        
			\legend{${n}=5$,
			        ${n}=20$,
			        ${n}=40$,
			        ${n}=\infty$
			        }           
		\end{loglogaxis}        
	\end{tikzpicture}
}

\newcommand{\FigAClossQfparamN}[0]{
    \begin{tikzpicture}
		\begin{loglogaxis}[	
            temflineplot,	
            legend pos  = north east,
            width       = 8cm, 
            height      = 6cm,
            xmin        = 1e-4, 
            xmax        = 1e+3,
            ymin        = 1e-5, 
            ymax        = 1e-3,	
            ytickten    = {-5,...,-3},
            xlabel      = ${f}$, 
            ylabel      = $\SCAl{w}{\RM{J}}$,
            x unit      = \si{\Hz}, 
            y unit      = \si{\joule\per\cubic\mm/cycle},
            ]
            
            \addplot+ 
                table[x=x,y=n5,col sep=comma] 
                {Data/ACloss_SIM_Q_f_B10mT_paramN.csv}; 
            \addplot+ 
                table[x=x,y=n20,col sep=comma] 
                {Data/ACloss_SIM_Q_f_B10mT_paramN.csv}; 
            \addplot+ 
                table[x=x,y=n40,col sep=comma] 
                {Data/ACloss_SIM_Q_f_B10mT_paramN.csv}; 
            \addplot+ 
                [mark=none, dashed, thin, gray!100] 
                table[x=xBrandt,y=nBrandt,col sep=comma] 
                {Data/ACloss_SIM_Q_f_B10mT_paramN.csv}; 
            \node 
                at (1e-2, 7e-4) 
                {$\SCAl{B}{\RM{p}}=\SI{10}{\milli\tesla}$};
            \node 
                at (2e-3, 1.7e-5) 
                {$\propto{f}^0$};
			
			\legend{${n}=5$,
			        ${n}=20$,
			        ${n}=40$,
			        ${n}=\infty$}              
		\end{loglogaxis}	
	\end{tikzpicture}
}

\newcommand{\FigAClossQIparamN}[0]{
    \begin{tikzpicture}
		\begin{loglogaxis}[	
            temflineplot,	
            legend pos  = north west,
            width       = 8cm, 
            height      = 6cm,
            xmin        = 1e0, 
            xmax        = 1e4,
            ymin        = 1e-11, 
            ymax        = 1e+1,
            ytickten    = {-11,...,+1},
            xlabel      = $\SCAl{I}{\RM{p}}$, 
            ylabel      = $\SCAl{w}{\RM{J}}$,
            x unit      = \si{\ampere}, 
            y unit      = \si{\joule\per\cubic\mm/cycle},
            ]
            
            \addplot+ 
                table[x=x,y=n5,col sep=comma]
                {Data/ACloss_SIM_Q_I_f1Hz_paramN.csv}; 
            \addplot+ 
                table[x=x,y=n20,col sep=comma, skip coords between index={10}{12}]
                {Data/ACloss_SIM_Q_I_f1Hz_paramN.csv}; 
            \addplot+ 
                table[x=x,y=n40,col sep=comma, skip coords between index={9}{12}]
                {Data/ACloss_SIM_Q_I_f1Hz_paramN.csv};             
            \addplot+ 
                [mark=none, dashed, thin, gray!100] 
                table[x=xBrandt,y=nBrandt,col sep=comma] 
                {Data/ACloss_SIM_Q_I_f1Hz_paramN.csv};     
            \node 
                at (4e1, 5e-10) 
                {$\propto\textrm{I}^4$};   
            \node 
                at (4e3, 7e-1) 
                {$\propto\textrm{I}^{{n}+1}$};       
            \addplot+ 
                [mark=none, dotted, thin, black]
                coordinates {(1e3,1e-12) (1e3,1e6)}; 
            \node 
                [black] 
                at (1.5e3, 1e-5) 
                {$\textrm{I}_\textrm{c}$};			
			\node 
			    [above right] 
			    at (rel axis cs:0.52,0.025) 
			    {\fcolorbox{black}{white}{\includegraphics[width=2.5cm]{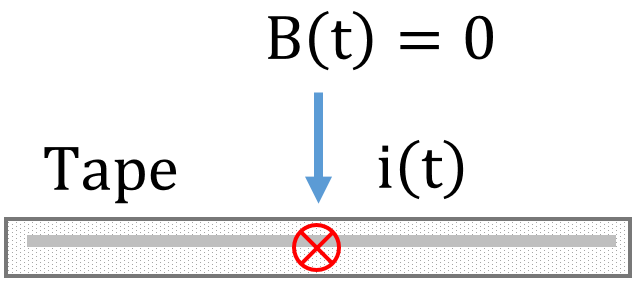}}}; 
			
			\legend{${n}=5$,
			        ${n}=20$,
			        ${n}=40$,
			        ${n}=\infty$,
			        }              
		\end{loglogaxis}	
	\end{tikzpicture}
}

\newcommand{\FigAClossQfparamI}[0]{
    \begin{tikzpicture}
		\begin{loglogaxis}[	
            temflineplot,
            legend pos  = north east,
            width       = 8cm, 
            height      = 6cm,
            xmin        = 1e-4, 
            xmax        = 1e3,
            ymin        = 1e-4, 
            ymax        = 1e-1,	
            ytickten    = {-4,...,-1},
            xlabel      = ${f}$, 
            ylabel      = $\SCAl{w}{\RM{J}}$,
            x unit      = \si{\Hz}, 
            y unit      = \si{\joule\per\cubic\mm/cycle},
            ]
            
            \addplot+ 
                table[x=x,y=n5,col sep=comma] 
                {Data/ACloss_SIM_Q_f_I500A_paramN.csv}; 
            \addplot+ 
                table[x=x,y=n20,col sep=comma]
                {Data/ACloss_SIM_Q_f_I500A_paramN.csv}; 
            \addplot+ 
                table[x=x,y=n40,col sep=comma]
                {Data/ACloss_SIM_Q_f_I500A_paramN.csv};             
            \addplot+ 
                [mark=none, dashed, thin, gray!100] 
                table[x=xBrandt,y=nBrandt,col sep=comma] 
                {Data/ACloss_SIM_Q_f_I500A_paramN.csv};     
            \node 
                at (1e-1, 5e-2) 
                {$\SCAl{I}{\RM{p}}=0.5\SCAl{I}{\RM{crit}}$};
            
            \node 
                at (1e-3, 2e-4) 
                {$\propto{f}^0$};
			
			\legend{${n}=5$,
			        ${n}=20$,
			        ${n}=40$,
			        ${n}=\infty$,
			        }             
	    \end{loglogaxis}
	\end{tikzpicture}
}

\newcommand{\FigureVerificationPlots}[0]{
    \begin{figure}[tb]
      \centering
        \FigAClossQBparamN
    	\caption{Specific Joule losses per cycle in a single HTS tape. Losses are given as a function of the sinusoidal magnetic field applied perpendicularly to the tape width, with peak value $\SCAl{B}{\RM{p}}$ and frequency $\SI{1}{Hz}$. The numerical results are parametrized by ${n}=5$, 20, 40, and compared with the analytical solution from literature where $n=\infty$.}
    	\label{FIG_AClossQBparamN}
        \vskip -0.25cm
    \end{figure}
    \begin{figure}[tb]
      \centering
        \FigAClossQfparamN
    	\caption{Specific Joule losses per cycle in a single HTS tape. Losses are given as a function of the sinusoidal magnetic field applied perpendicularly to the tape width, with peak value $\SCAl{B}{\RM{p}}=\SI{10}{\milli\tesla}$ and frequency $f$. The numerical results are parametrized by ${n}=5$, 20, 40, and compared with the analytical solution from literature where $n=\infty$.}
    	\label{FIG_AClossQfparamN}
        \vskip -0.25cm
    \end{figure}
    \begin{figure}[tb]
      \centering
        \FigAClossQIparamN
    	\caption{Specific Joule losses per cycle in a single HTS tape of critical current $\SCAl{I}{\RM{c}}=1\si{\kilo\ampere}$. Losses are given as a function of the sinusoidal supply current, with peak value $\SCAl{I}{\RM{p}}$ and frequency $\SI{1}{Hz}$. The numerical results are parametrized by ${n}=5$, 20, 40, and compared with the analytical solution from literature where $n=\infty$.}
    	\label{FIG_AClossQIparamN}
        \vskip -0.25cm
    \end{figure}
    \begin{figure}[tb]
      \centering
        \FigAClossQfparamI
    	\caption{Specific Joule losses per cycle in a single HTS tape of critical current $\SCAl{I}{\RM{c}}=1\si{\kilo\ampere}$. Losses are given as a function of the sinusoidal supply current, with peak value $\SCAl{I}{\RM{p}}=0.5\SCAl{I}{\RM{c}}$ and frequency $f$. The numerical results are parametrized by ${n}=5$, 20, 40, and compared with the analytical solution from literature where $n=\infty$.}
    	\label{FIG_AClossQfparamI}
        \vskip -0.25cm
    \end{figure}
}

%% file: PGFplot/FIG_Result.tex
\newcommand{\plotAbsMutipole}[2]{
            
    \addplot 
        [ultra thin, only marks, 
        mark options={mark = +, mark size=2pt,black}] 
        table[x=I(kA)_up,y=#1_up,col sep=comma, dashed] 
        {Data/B_MEAS_#2.csv};   
    \addplot 
        [ultra thin, only marks, 
        mark options={solid,mark = x, mark size=2pt,black}] 
        table[x=I(kA)_dn,y=#1_dn,col sep=comma, dashed] 
        {Data/B_MEAS_#2.csv};   
    
	\addplot
		[name path=A,draw=none,forget plot] 
		table[x=I(kA),y=#1_min,col sep=comma] 
        {Data/B_SIM_#2_MINMAX.csv};                   
	\addplot
		[name path=B,draw=none,forget plot] 
		table[x=I(kA),y=#1_max,col sep=comma] 
        {Data/B_SIM_#2_MINMAX.csv};                 
    \addplot
        [gray!75,fill opacity=0.45] 
        fill between[of=A and B];  
        
    \pgfplotsset{cycle list shift=-3} 
    \addplot+ 
        [no markers] 
        table[x=I(kA),y=#1,col sep=comma] 
        {Data/B_SIM_#2.csv}; 

    \addplot
        [mark=none, dashed, blue] 
        table[x=I(kA),y=#1,col sep=comma] 
        {Data/B_SIM_#2_homogeneousJ.csv};              
}

\newcommand{\FigAbsBoneFirst}[0]{
    	\begin{tikzpicture}
    	\def\multipoleName{B1}  
    	\def\TopFileName{4p5K} 	
		\begin{axis}[	
            temflineplot,
            title       = {$\SCAl{T}{\RM{op}}=\SI{4.5}{\kelvin}$},
            width       = \figWidth,
            height      = \figHeight,
            xmin        = 0, 
            xmax        = 5,
            xtick       = {0,...,5},
            ymin        = 0, 
            ymax        = 3,
            ytick       = {0,...,3},
            ylabel      = $\mathrm{B_1}$,
            y unit      = \si{\tesla},
            legend style   = {font=\normalsize,legend columns=-1,fill=none,draw=black,anchor=center},
            legend entries = {up$\ \ \ \ \ $,down$\ \ \ \ \ $,n=sweep$\ \ \ \ \ $,n=20$\ \ \ \ \ $,$|\VEC{J}|$ homogeneous$\ \ \ \ \ $},
            legend to name = myLegendBabs
            ]            
            \plotAbsMutipole{\multipoleName}{\TopFileName}
		\end{axis}
	\end{tikzpicture}
}

\newcommand{\FigAbsBoneSecond}[0]{
    	\begin{tikzpicture}
    	\def\TopFileName{9p0K}
     	\def\multipoleName{B1}   	
		\begin{axis}[	
            temflineplot,	    
            title       = {$\SCAl{T}{\RM{op}}=\SI{9}{\kelvin}$},
            width       = \figWidth, 
            height      = \figHeight,
            xmin        = 0, 
            xmax        = 5,
            xtick       = {0,...,5},
            ymin        = 0, 
            ymax        = 3,
            ytick       = {0,...,3},
            ]         
            \plotAbsMutipole{\multipoleName}{\TopFileName}            
		\end{axis}
	\end{tikzpicture}
}

\newcommand{\FigAbsBoneThird}[0]{
    	\begin{tikzpicture}
    	\def\TopFileName{25p0K}
     	\def\multipoleName{B1}   	
		\begin{axis}[	
            temflineplot,	    
            title       = {$\SCAl{T}{\RM{op}}=\SI{25}{\kelvin}$},
            width       = \figWidth, 
            height      = \figHeight,
            xmin        = 0, 
            xmax        = 5,
            xtick       = {0,...,5},
            ymin        = 0, 
            ymax        = 3,
            ytick       = {0,...,3},	
            ]               
            \plotAbsMutipole{\multipoleName}{\TopFileName}            
		\end{axis}
	\end{tikzpicture}
}

\newcommand{\FigAbsBoneFifth}[0]{
    	\begin{tikzpicture}
    	\def\TopFileName{68p0K}
     	\def\multipoleName{B1}   	
		\begin{axis}[	
            temflineplot,	    
            title       = {$\RM{{T}_{op}}=\SI{68}{\kelvin}$},
            width       = \figWidth, 
            height      = \figHeight,
            xmin        = 0, 
            xmax        = 5,
            xtick       = {0,...,5},
            ymin        = 0, 
            ymax        = 3,
            ytick       = {0,...,3},	
            ]                                  
            \plotAbsMutipole{\multipoleName}{\TopFileName}            
		\end{axis}
	\end{tikzpicture}
}

\newcommand{\FigAbsBthreeFirst}[0]{
    	\begin{tikzpicture}
    	\def\TopFileName{4p5K}
     	\def\multipoleName{b3}   	
		\begin{axis}[	
            temflineplot,	    
            width       = \figWidth, 
            height      = \figHeight,
            xmin        = 0, 
            xmax        = 5,
            xtick       = {0,...,5},
            ymin        = 0, 
            ymax        = 500,
            ytick       = {0,100,...,500},
            ylabel      = $\mathrm{b_3}$,
            y unit      = \si{units},               
            ]                                  
            \plotAbsMutipole{\multipoleName}{\TopFileName}            
		\end{axis}
	\end{tikzpicture}
}

\newcommand{\FigAbsBthreeSecond}[0]{
    	\begin{tikzpicture}
    	\def\TopFileName{9p0K}
     	\def\multipoleName{b3}   	
		\begin{axis}[	
            temflineplot,	    
            width       = \figWidth, 
            height      = \figHeight,
            xmin        = 0, 
            xmax        = 5,
            xtick       = {0,...,5},
            ymin        = 0, 
            ymax        = 500,
            ytick       = {0,100,...,500},
            ]                                 
            \plotAbsMutipole{\multipoleName}{\TopFileName}            
		\end{axis}
	\end{tikzpicture}
}

\newcommand{\FigAbsBthreeThird}[0]{
    	\begin{tikzpicture}
    	\def\TopFileName{25p0K}
     	\def\multipoleName{b3}   	
		\begin{axis}[	
            temflineplot,	    
            width       = \figWidth, 
            height      = \figHeight,
            xmin        = 0, 
            xmax        = 5,
            xtick       = {0,...,5},
            ymin        = 0, 
            ymax        = 500,
            ytick       = {0,100,...,500},
            ]                                 
            \plotAbsMutipole{\multipoleName}{\TopFileName}            
		\end{axis}
	\end{tikzpicture}
}

\newcommand{\FigAbsBthreeFifth}[0]{
    	\begin{tikzpicture}
    	\def\TopFileName{68p0K}
     	\def\multipoleName{b3}   	
		\begin{axis}[	
            temflineplot,	    
            width       = \figWidth, 
            height      = \figHeight,
            xmin        = 0, 
            xmax        = 5,
            xtick       = {0,...,5},
            ymin        = 0, 
            ymax        = 500,
            ytick       = {0,100,...,500},
            ]                                  
            \plotAbsMutipole{\multipoleName}{\TopFileName}            
		\end{axis}
	\end{tikzpicture}
}

\newcommand{\FigAbsBfiveFirst}[0]{
    	\begin{tikzpicture}
    	\def\TopFileName{4p5K}
     	\def\multipoleName{b5}   	
		\begin{axis}[	
            temflineplot,	    
            width       = \figWidth, 
            height      = \figHeight,
            xmin        = 0, 
            xmax        = 5,
            xtick       = {0,...,5},
            ymin        = 0, 
            ymax        = 100,	
            ylabel      = $\mathrm{b_5}$,
            y unit      = \si{units},               
            ]                                  
            \plotAbsMutipole{\multipoleName}{\TopFileName}            
		\end{axis}
	\end{tikzpicture}
}

\newcommand{\FigAbsBfiveSecond}[0]{
    	\begin{tikzpicture}
    	\def\TopFileName{9p0K}
     	\def\multipoleName{b5}   	
		\begin{axis}[	
            temflineplot,	    
            width       = \figWidth, 
            height      = \figHeight,
            xmin        = 0, 
            xmax        = 5,
            xtick       = {0,...,5},
            ymin        = 0, 
            ymax        = 100,	
            ]                                 
            \plotAbsMutipole{\multipoleName}{\TopFileName}            
		\end{axis}
	\end{tikzpicture}
}

\newcommand{\FigAbsBfiveThird}[0]{
    	\begin{tikzpicture}
    	\def\TopFileName{25p0K}
     	\def\multipoleName{b5}   	
		\begin{axis}[	
            temflineplot,	    
            width       = \figWidth, 
            height      = \figHeight,
            xmin        = 0, 
            xmax        = 5,
            xtick       = {0,...,5},
            ymin        = 0, 
            ymax        = 100,	
            ]                                  
            \plotAbsMutipole{\multipoleName}{\TopFileName}            
		\end{axis}
	\end{tikzpicture}
}

\newcommand{\FigAbsBfiveFifth}[0]{
    	\begin{tikzpicture}
    	\def\TopFileName{68p0K}
     	\def\multipoleName{b5}   	
		\begin{axis}[	
            temflineplot,	    
            width       = \figWidth, 
            height      = \figHeight,
            xmin        = 0, 
            xmax        = 5,
            xtick       = {0,...,5},
            ymin        = 0, 
            ymax        = 100,	
            ]                                  
            \plotAbsMutipole{\multipoleName}{\TopFileName}            
		\end{axis}
	\end{tikzpicture}
}

\newcommand{\FigAbsBsevenFirst}[0]{
    	\begin{tikzpicture}
    	\def\TopFileName{4p5K}
     	\def\multipoleName{b7}   	
		\begin{axis}[	
            temflineplot,	    
            width       = \figWidth, 
            height      = \figHeight,
            xmin        = 0, 
            xmax        = 5,
            xtick       = {0,...,5},
            ymin        = 0, 
            ymax        = 10,	
            xlabel      = , 
            ylabel      = $\mathrm{b_7}$,
            x unit      = \si{\kilo\ampere},  
            y unit      = \si{units},               
            ]                                 
            \plotAbsMutipole{\multipoleName}{\TopFileName}            
		\end{axis}
	\end{tikzpicture}
}

\newcommand{\FigAbsBsevenSecond}[0]{
    	\begin{tikzpicture}
    	\def\TopFileName{9p0K}
     	\def\multipoleName{b7}   	
		\begin{axis}[	
            temflineplot,	    
            width       = \figWidth, 
            height      = \figHeight,
            xmin        = 0, 
            xmax        = 5,
            xtick       = {0,...,5},
            ymin        = 0, 
            ymax        = 10,	
            xlabel      = , 
            x unit      = \si{\kilo\ampere},  
            ]                                  
            \plotAbsMutipole{\multipoleName}{\TopFileName}            
		\end{axis}
	\end{tikzpicture}
}

\newcommand{\FigAbsBsevenThird}[0]{
    	\begin{tikzpicture}
    	\def\TopFileName{25p0K}
     	\def\multipoleName{b7}   	
		\begin{axis}[	
            temflineplot,	    
            width       = \figWidth, 
            height      = \figHeight,
            xmin        = 0, 
            xmax        = 5,
            xtick       = {0,...,5},
            ymin        = 0, 
            ymax        = 10,	
            xlabel      = , 
            x unit      = \si{\kilo\ampere},  
            ]                                  
            \plotAbsMutipole{\multipoleName}{\TopFileName}            
		\end{axis}
	\end{tikzpicture}
}

\newcommand{\FigAbsBsevenFifth}[0]{
    	\begin{tikzpicture}
    	\def\TopFileName{68p0K}
     	\def\multipoleName{b7}   	
		\begin{axis}[	
            temflineplot,	    
            width       = \figWidth, 
            height      = \figHeight,
            xmin        = 0, 
            xmax        = 5,
            xtick       = {0,...,5},
            ymin        = 0, 
            ymax        = 10,	
            xlabel      = , 
            x unit      = \si{\kilo\ampere},  
            ]                                  
            \plotAbsMutipole{\multipoleName}{\TopFileName}            
		\end{axis}
	\end{tikzpicture}
}

 \newcommand{\plotRelMutipole}[2]{
            
    \addplot 
        [ultra thin, only marks, 
        mark options={mark = triangle*, mark size=2pt,black}]  
        table[x=I_delta(kA),y=delta_#1,col sep=comma, dashed] 
        {Data/B_MEAS_#2.csv};   

	\addplot
		[name path=A,draw=none,forget plot] 
		table[x=I(kA),y=delta_#1_min,col sep=comma] 
        {Data/B_SIM_#2_MINMAX.csv};                   
	\addplot
		[name path=B,draw=none,forget plot] 
	    table[x=I(kA),y=delta_#1_max,col sep=comma] 
        {Data/B_SIM_#2_MINMAX.csv};                 
    \addplot
        [gray!75,fill opacity=0.45] 
        fill between[of=A and B];
        
    \pgfplotsset{cycle list shift=-2} 
    \addplot+ 
        [no markers] 
        table[col sep=comma,
            x=I_delta(kA),
            y={delta_#1},
            ] 
        {Data/B_SIM_#2.csv};  
        
    \addplot
        [mark=none, dashed, blue] 
        table[x=I_delta(kA),y={delta_#1},col sep=comma] 
        {Data/B_SIM_#2_homogeneousJ.csv};  
}

\newcommand{\FigRelBoneFirst}[0]{
    	\begin{tikzpicture}
    	\def\TopFileName{4p5K}
     	\def\multipoleName{b1}   	
		\begin{axis}[	
            temflineplot,	    
            title       = {$\SCAl{T}{\RM{op}}=\SI{4.5}{\kelvin}$},
            width       = \figWidth, 
            height      = \figHeight,
            xmin        = 0, 
            xmax        = 5,
            ymin        = -20, 
            ymax        = 60,
            xtick       = {0,...,5},
            ylabel      = $\mathrm{b_1}$,
            y unit      = \si{units},
            legend style   = {font=\normalsize,legend columns=-1,fill=none,draw=black,anchor=center},
            legend entries = {meas.$\ \ \ \ \ $,n=sweep$\ \ \ \ \ $,n=20$\ \ \ \ \ $,$|\VEC{J}|$ homogeneous$\ \ \ \ \ $},
            legend to name = myLegendBrel                     
            ]            
            \plotRelMutipole{\multipoleName}{\TopFileName}             			
		\end{axis}
	\end{tikzpicture}
}

\newcommand{\FigRelBoneSecond}[0]{
    	\def\TopFileName{9p0K}
     	\def\multipoleName{b1}   	
    	\begin{tikzpicture}
		\begin{axis}[	
            temflineplot,	    
            title       = {$\SCAl{T}{\RM{op}}=\SI{9}{\kelvin}$},
            width       = \figWidth, 
            height      = \figHeight,
            xmin        = 0, 
            xmax        = 5,
            ymin        = -20, 
            ymax        = 60,
            xtick       = {0,...,5},
            ]                                
            \plotRelMutipole{\multipoleName}{\TopFileName}             
		\end{axis}
	\end{tikzpicture}
}

\newcommand{\FigRelBoneThird}[0]{
    	\begin{tikzpicture}
    	\def\TopFileName{25p0K}
     	\def\multipoleName{b1}   	
		\begin{axis}[	
            temflineplot,	    
            title       = {$\SCAl{T}{\RM{op}}=\SI{25}{\kelvin}$},
            width       = \figWidth, 
            height      = \figHeight,
            xmin        = 0, 
            xmax        = 5,
            ymin        = -20, 
            ymax        = 60,
            xtick       = {0,...,5},
            ]                                  
            \plotRelMutipole{\multipoleName}{\TopFileName}             
		\end{axis}
	\end{tikzpicture}
}

\newcommand{\FigRelBoneFifth}[0]{
    	\begin{tikzpicture}
    	\def\TopFileName{68p0K}
     	\def\multipoleName{b1}   	
		\begin{axis}[	
            temflineplot,	    
            title       = {$\SCAl{T}{\RM{op}}=\SI{68}{\kelvin}$},
            width       = \figWidth, 
            height      = \figHeight,
            xmin        = 0, 
            xmax        = 5,
            ymin        = -20, 
            ymax        = 60,
            xtick       = {0,...,5},
            ]                                  
            \plotRelMutipole{\multipoleName}{\TopFileName}            
		\end{axis}
	\end{tikzpicture}
}

\newcommand{\FigRelBthreeFirst}[0]{
    	\begin{tikzpicture}
    	\def\TopFileName{4p5K}
     	\def\multipoleName{b3}   	
		\begin{axis}[	
            temflineplot,	    
            width       = \figWidth, 
            height      = \figHeight,
            xmin        = 0, 
            xmax        = 5,
            ymin        = -20, 
            ymax        = 40,
            xtick       = {0,...,5},
            ylabel      = $\mathrm{b_3}$,
            y unit      = \si{units},               
            ]                                  
            \plotRelMutipole{\multipoleName}{\TopFileName}             
		\end{axis}
	\end{tikzpicture}
}

\newcommand{\FigRelBthreeSecond}[0]{
    	\begin{tikzpicture}
    	\def\TopFileName{9p0K}
     	\def\multipoleName{b3}   	
		\begin{axis}[	
            temflineplot,	    
            width       = \figWidth, 
            height      = \figHeight,
            xmin        = 0, 
            xmax        = 5,
            ymin        = -20, 
            ymax        = 40,
            xtick       = {0,...,5},
            ]                                  
            \plotRelMutipole{\multipoleName}{\TopFileName}             
		\end{axis}
	\end{tikzpicture}
}

\newcommand{\FigRelBthreeThird}[0]{
    	\begin{tikzpicture}
    	\def\TopFileName{25p0K}
     	\def\multipoleName{b3}   	
		\begin{axis}[	
            temflineplot,	    
            width       = \figWidth, 
            height      = \figHeight,
            xmin        = 0, 
            xmax        = 5,
            ymin        = -20, 
            ymax        = 40,
            xtick       = {0,...,5},
            ]                                  
            \plotRelMutipole{\multipoleName}{\TopFileName}
		\end{axis}
	\end{tikzpicture}
}

\newcommand{\FigRelBthreeFifth}[0]{
    	\begin{tikzpicture}
    	\def\TopFileName{68p0K}
     	\def\multipoleName{b3}   	
		\begin{axis}[	
            temflineplot,	    
            width       = \figWidth, 
            height      = \figHeight,
            xmin        = 0, 
            xmax        = 5,
            ymin        = -20, 
            ymax        = 40,
            xtick       = {0,...,5},
            ]                                 
            \plotRelMutipole{\multipoleName}{\TopFileName}             
		\end{axis}
	\end{tikzpicture}
}

\newcommand{\FigRelBfiveFirst}[0]{
    	\begin{tikzpicture}
    	\def\TopFileName{4p5K}
     	\def\multipoleName{b5}   	
		\begin{axis}[	
            temflineplot,	    
            width       = \figWidth, 
            height      = \figHeight,
            xmin        = 0, 
            xmax        = 5,
            ymin        = -4, 
            ymax        = 8,
            xtick       = {0,...,5},
            ytick       = {-4,0,...,8},
            ylabel      = $\mathrm{b_5}$,
            y unit      = \si{units},               
            ]                                  
            \plotRelMutipole{\multipoleName}{\TopFileName}             
		\end{axis}
	\end{tikzpicture}
}

\newcommand{\FigRelBfiveSecond}[0]{
    	\begin{tikzpicture}
    	\def\TopFileName{9p0K}
     	\def\multipoleName{b5}   	
		\begin{axis}[	
            temflineplot,	    
            width       = \figWidth, 
            height      = \figHeight,
            xmin        = 0, 
            xmax        = 5,
            ymin        = -4, 
            ymax        = 8,
            xtick       = {0,...,5},
            ytick       = {-4,0,...,8},
            ]                                  
            \plotRelMutipole{\multipoleName}{\TopFileName}             
		\end{axis}
	\end{tikzpicture}
}

\newcommand{\FigRelBfiveThird}[0]{
    	\begin{tikzpicture}
    	\def\TopFileName{25p0K}
     	\def\multipoleName{b5}   	
		\begin{axis}[	
            temflineplot,	    
            width       = \figWidth, 
            height      = \figHeight,
            xmin        = 0, 
            xmax        = 5,
            ymin        = -4, 
            ymax        = 8,
            xtick       = {0,...,5},
            ytick       = {-4,0,...,8},
            ]                                  
            \plotRelMutipole{\multipoleName}{\TopFileName}             
		\end{axis}
	\end{tikzpicture}
}

\newcommand{\FigRelBfiveFifth}[0]{
    	\begin{tikzpicture}
    	\def\TopFileName{68p0K}
     	\def\multipoleName{b5}   	
		\begin{axis}[	
            temflineplot,	    
            width       = \figWidth, 
            height      = \figHeight,
            xmin        = 0, 
            xmax        = 5,
            ymin        = -4, 
            ymax        = 8,
            xtick       = {0,...,5},
            ytick       = {-4,0,...,8},
            ]                                 
            \plotRelMutipole{\multipoleName}{\TopFileName}             
		\end{axis}
	\end{tikzpicture}
}

\newcommand{\FigRelBsevenFirst}[0]{
    	\begin{tikzpicture}
    	\def\TopFileName{4p5K}
     	\def\multipoleName{b7}   	
		\begin{axis}[	
            temflineplot,	    
            width       = \figWidth, 
            height      = \figHeight,
            xmin        = 0, 
            xmax        = 5,
            ymin        = -4, 
            ymax        = 4,
            xtick       = {0,...,5},
            ytick       = {-4,-2,...,4},
            xlabel      = , 
            ylabel      = $\mathrm{b_7}$,
            x unit      = \si{\kilo\ampere}, 
            y unit      = \si{units},               
            ]                                  
            \plotRelMutipole{\multipoleName}{\TopFileName}             
		\end{axis}
	\end{tikzpicture}
}

\newcommand{\FigRelBsevenSecond}[0]{
    	\begin{tikzpicture}
    	\def\TopFileName{9p0K}
     	\def\multipoleName{b7}   	
		\begin{axis}[	
            temflineplot,	    
            width       = \figWidth, 
            height      = \figHeight,
            xmin        = 0, 
            xmax        = 5,
            ymin        = -4, 
            ymax        = 4,
            xtick       = {0,...,5},
            ytick       = {-4,-2,...,4},
            xlabel      = , 
            x unit      = \si{\kilo\ampere}, 
            ]                                  
            \plotRelMutipole{\multipoleName}{\TopFileName}             
		\end{axis}
	\end{tikzpicture}
}

\newcommand{\FigRelBsevenThird}[0]{
    	\begin{tikzpicture}
    	\def\TopFileName{25p0K}
     	\def\multipoleName{b7}   	
		\begin{axis}[	
            temflineplot,	    
            width       = \figWidth, 
            height      = \figHeight,
            xmin        = 0, 
            xmax        = 5,
            ymin        = -4, 
            ymax        = 4,
            xtick       = {0,...,5},
            ytick       = {-4,-2,...,4},
            xlabel      = , 
            x unit      = \si{\kilo\ampere}, 
            ]                                 
            \plotRelMutipole{\multipoleName}{\TopFileName}             
		\end{axis}
	\end{tikzpicture}
}

\newcommand{\FigRelBsevenFifth}[0]{
    	\begin{tikzpicture}
    	\def\TopFileName{68p0K}
     	\def\multipoleName{b7}   	
		\begin{axis}[	
            temflineplot,	    
            width       = \figWidth, 
            height      = \figHeight,
            xmin        = 0, 
            xmax        = 5,
            ymin        = -4, 
            ymax        = 4,
            xtick       = {0,...,5},
            ytick       = {-4,-2,...,4},
            xlabel      = , 
            x unit      = \si{\kilo\ampere}, 
            ]                                
            \plotRelMutipole{\multipoleName}{\TopFileName}             
		\end{axis}
	\end{tikzpicture}
}

\newcommand{\FigureResultPlots}[0]{
\begin{figure*}[tb]
\raggedright
    \begin{minipage}[l]{\MyClassDependentFigResultSpan}
        \raggedright 
            \hspace*{2.0mm}
            \FigAbsBoneFirst
        \vskip -0.35cm
    \end{minipage}
    \hfill{}
    \begin{minipage}[l]{\MyClassDependentFigResultSpan}
        \raggedright 
          \hspace*{2.0mm}
          \FigAbsBoneSecond
        \vskip -0.35cm
    \end{minipage}
    \hfill{}
    \begin{minipage}[l]{\MyClassDependentFigResultSpan}
        \raggedright 
            \hspace*{2.0mm}
            \FigAbsBoneThird
        \vskip -0.35cm
    \end{minipage}
    \hfill{}
    \begin{minipage}[l]{\MyClassDependentFigResultSpan}
        \raggedright 
            \hspace*{2.0mm}
            \FigAbsBoneFifth
        \vskip -0.35cm
    \end{minipage}
    \\
    \begin{minipage}[l]{\MyClassDependentFigResultSpan}
        \raggedright
            \FigAbsBthreeFirst
        \vskip -0.35cm
    \end{minipage}
    \hfill{}
    \begin{minipage}[l]{\MyClassDependentFigResultSpan}
        \raggedright 
            \FigAbsBthreeSecond
        \vskip -0.35cm
    \end{minipage}
    \hfill{}
    \begin{minipage}[l]{\MyClassDependentFigResultSpan}
        \raggedright
            \FigAbsBthreeThird
        \vskip -0.35cm
    \end{minipage}
    \hfill{}
    \begin{minipage}[l]{\MyClassDependentFigResultSpan}
        \raggedright
            \FigAbsBthreeFifth
        \vskip -0.35cm
    \end{minipage}
    \\
    \begin{minipage}[l]{\MyClassDependentFigResultSpan}
        \raggedright 
            \FigAbsBfiveFirst
        \vskip -0.35cm
    \end{minipage}
    \hfill{}
    \begin{minipage}[l]{\MyClassDependentFigResultSpan}
        \raggedright 
            \FigAbsBfiveSecond
        \vskip -0.35cm
    \end{minipage}
    \hfill{}
    \begin{minipage}[l]{\MyClassDependentFigResultSpan}
        \raggedright 
            \FigAbsBfiveThird
        \vskip -0.35cm
    \end{minipage}
    \hfill{}
    \begin{minipage}[l]{\MyClassDependentFigResultSpan}
        \raggedright 
            \FigAbsBfiveFifth
        \vskip -0.35cm
    \end{minipage}
        \\
    \begin{minipage}[l]{\MyClassDependentFigResultSpan}
        \raggedright 
            \hspace*{1.0mm}\FigAbsBsevenFirst
        \vskip -0.35cm
    \end{minipage}
    \hfill{}
    \begin{minipage}[l]{\MyClassDependentFigResultSpan}
        \raggedright 
            \hspace*{1.0mm}\FigAbsBsevenSecond
        \vskip -0.35cm
    \end{minipage}
    \hfill{}
    \begin{minipage}[l]{\MyClassDependentFigResultSpan}
        \raggedright 
            \hspace*{1.0mm}\FigAbsBsevenThird
        \vskip -0.35cm
    \end{minipage}
    \hfill{}
    \begin{minipage}[l]{\MyClassDependentFigResultSpan}
        \raggedright 
            \hspace*{1.0mm}\FigAbsBsevenFifth
        \vskip -0.35cm
    \end{minipage}
    \\
    \centering
    \ref{myLegendBabs}
    \caption{   
    Magnetic field quality in the magnet aperture as a function of the current, using a current staircase profile (see Fig.~\ref{FIG_NormalizedStaircase}). Measurements are given by markers, whereas the shaded area corresponds to the envelope of the numerical solutions, obtained with the parametric sweep of the $n$-value as $4\leq{n}\leq{30}$. The solution for $n=20$ is marked with a solid line. The dotted line is obtained by assuming a homogeneous current density distribution in the superconducting tapes. From left to right: results at 4.5, 9, 25, and $\SI{68}{\kelvin}$. From top to bottom: results for the $\SCAl{B}{1}$, $\SCAl{b}{3}$, $\SCAl{b}{5}$, $\SCAl{b}{7}$ multipole coefficients.
    }
	\label{FIG_FieldQuality}
\end{figure*}

\begin{figure*}[tb]
\raggedright
    \begin{minipage}[l]{\MyClassDependentFigResultSpan}
        \raggedright 
            \FigRelBoneFirst
        \vskip -0.35cm
    \end{minipage}
    \hfill{}
    \begin{minipage}[l]{\MyClassDependentFigResultSpan}
        \raggedright 
            \FigRelBoneSecond
        \vskip -0.35cm
    \end{minipage}
    \hfill{}
    \begin{minipage}[l]{\MyClassDependentFigResultSpan}
        \raggedright 
            \FigRelBoneThird
        \vskip -0.35cm
    \end{minipage}
    \hfill{}
    \begin{minipage}[l]{\MyClassDependentFigResultSpan}
        \raggedright 
            \FigRelBoneFifth
        \vskip -0.35cm
    \end{minipage}
    \\
    \begin{minipage}[l]{\MyClassDependentFigResultSpan}
        \raggedright 
            \FigRelBthreeFirst
        \vskip -0.35cm
    \end{minipage}
    \hfill{}
    \begin{minipage}[l]{\MyClassDependentFigResultSpan}
        \raggedright 
            \FigRelBthreeSecond
        \vskip -0.35cm
    \end{minipage}
    \hfill{}
    \begin{minipage}[l]{\MyClassDependentFigResultSpan}
        \raggedright 
            \FigRelBthreeThird
        \vskip -0.35cm
    \end{minipage}
    \hfill{}
    \begin{minipage}[l]{\MyClassDependentFigResultSpan}
        \raggedright 
            \FigRelBthreeFifth
        \vskip -0.35cm
    \end{minipage}
    \\
    \begin{minipage}[l]{\MyClassDependentFigResultSpan}
        \raggedright
            \hspace*{0.3mm}
            \FigRelBfiveFirst
        \vskip -0.35cm
    \end{minipage}
    \hfill{}
    \begin{minipage}[l]{\MyClassDependentFigResultSpan}
        \raggedright 
            \hspace*{0.3mm}
            \FigRelBfiveSecond
        \vskip -0.35cm
    \end{minipage}
    \hfill{}
    \begin{minipage}[l]{\MyClassDependentFigResultSpan}
        \raggedright
            \hspace*{0.3mm}
            \FigRelBfiveThird
        \vskip -0.35cm
    \end{minipage}
    \hfill{}
    \begin{minipage}[l]{\MyClassDependentFigResultSpan}
        \raggedright
            \hspace*{0.3mm}
            \FigRelBfiveFifth
        \vskip -0.35cm
    \end{minipage}
        \\
    \begin{minipage}[l]{\MyClassDependentFigResultSpan}
        \raggedright
            \hspace*{0.3mm}
            \FigRelBsevenFirst
        \vskip -0.35cm
    \end{minipage}
    \hfill{}
    \begin{minipage}[l]{\MyClassDependentFigResultSpan}
        \raggedright
            \hspace*{0.3mm}
            \FigRelBsevenSecond
        \vskip -0.35cm
    \end{minipage}
    \hfill{}
    \begin{minipage}[l]{\MyClassDependentFigResultSpan}
        \raggedright
            \hspace*{0.3mm}
            \FigRelBsevenThird
        \vskip -0.35cm
    \end{minipage}
    \hfill{}
    \begin{minipage}[l]{\MyClassDependentFigResultSpan}
        \raggedright
            \hspace*{0.3mm}
            \FigRelBsevenFifth
        \vskip -0.35cm
    \end{minipage}
    \\
    \centering
    \ref{myLegendBrel}
    \caption{
    Screening currents-induced magnetic field contribution to the magnetic field quality, in units, as a function of the current in the magnet. Measurements are given by markers, whereas the shaded area corresponds to the envelope of the numerical solutions, obtained with the parametric sweep of the $n$-value as $4\leq{n}\leq{30}$. The solution for $n=20$ is marked with a solid line. The dotted line is obtained by assuming a homogeneous current density distribution in the superconducting tapes. From left to right: results at 4.5, 9, 25, and $\SI{68}{\kelvin}$. From top to bottom: results for the $\SCAl{b}{1}$, $\SCAl{b}{3}$, $\SCAl{b}{5}$, $\SCAl{b}{7}$ multipole coefficients.
    }
	\label{FIG_FieldQualityPersistentCurrent}
\end{figure*}
}

%% file: PGFplot/FIG_Other.tex
\newcommand{\FigMeshSensitivity}[0]{
    	\begin{tikzpicture}
		\begin{loglogaxis}[	
            temflineplot,
            legend pos  = north east,
            width       = 8cm, 
            height      = 6cm,
            xmin        = 1e1, 
            xmax        = 1e4,
            ymin        = 1e-5, 
            ymax        = 1e2,
            ytickten    = {-5,...,2},
            xlabel      = $\frac{1}{\SCA{\RMDelta{x}}}$, 
            ylabel      = $\SCAl{\RMepsilon}{\SCA{\RMDelta{x}}}$,
            x unit      = \si{\per\meter}, 
            y unit      = \si{-},       
            ]            
            
            \addplot+ 
                table[x=meshDimInv,y=errTHDrel,col sep=comma] 
                {Data/MeshSensitivity.csv}; \label{plot_one}		
		\end{loglogaxis}
		
		\begin{loglogaxis}[	
            temflineplot,
		    axis y line*= right,
		    xmajorgrids = false,
	    	ymajorgrids = false,
            legend pos  = north east,
            width       = 8cm, 
            height      = 6cm,
            xmin        = 1e1, 
            xmax        = 1e4,
            ymin        = 1e2, 
            ymax        = 1e6,
            xlabel      = , 
            ylabel      = no. of elements,
            x unit      = , 
            y unit      = \si{-}, 
            hide x axis
            ]  
            
            \addlegendimage{/pgfplots/refstyle=plot_one}\addlegendentry{THD error} 
            
            \pgfplotsset{cycle list shift=+1}
            \addplot+ 
                table[x=meshDimInv,y=noOfElem,col sep=comma] 
                {Data/MeshSensitivity.csv}; \label{plot_two}
                
            \addlegendimage{/pgfplots/refstyle=plot_two}\addlegendentry{no. of elements} 
      	
		\end{loglogaxis}
				
	\end{tikzpicture}
}

\newcommand{\FigIcFitFunB}[0]{
    	\begin{tikzpicture}
		\begin{axis}[	
            temflineplot,
            width       = 8cm, 
            height      = 6cm,
            xmin        = 0, 
            xmax        = 8,
            xtick       = {0,1,...,8},
            ymin        = 0, 
            ymax        = 12,
            ytick       = {0,2,...,12},
            xlabel      = $\SCA{B}$, 
            ylabel      = $\SCAl{I}{\RM{c}}$,
            x unit      = \si{\tesla}, 
            y unit      = \si{\kilo\ampere},       
            ]            
            
            \addplot+ 
                [no markers] 
                table[x=B(T),y=T10K,col sep=comma] 
                {Data/Ic_Fit.csv}; 
            \addplot+ 
                [no markers] 
                table[x=B(T),y=T20K,col sep=comma] 
                {Data/Ic_Fit.csv}; 
            \addplot+ 
                [no markers] 
                table[x=B(T),y=T30K,col sep=comma] 
                {Data/Ic_Fit.csv}; 
            \addplot+ 
                [no markers] 
                table[x=B(T),y=T40K,col sep=comma] 
                {Data/Ic_Fit.csv}; 
            \addplot+ 
                [no markers] 
                table[x=B(T),y=T50K,col sep=comma] 
                {Data/Ic_Fit.csv}; 
            \addplot+ 
                [no markers] 
                table[x=B(T),y=T60K,col sep=comma] 
                {Data/Ic_Fit.csv}; 
            \addplot+ 
                [no markers] 
                table[x=B(T),y=T70K,col sep=comma] 
                {Data/Ic_Fit.csv}; 
            \addplot+ 
                [no markers] 
                table[x=B(T),y=T80K,col sep=comma] 
                {Data/Ic_Fit.csv};  
            \addplot+ 
                [dotted, black]
                [no markers] 
                table[x=Bload(T),y=Iload(kA),col sep=comma, dashed] 
                {Data/Ic_Fit.csv};  
            \addplot+ 
                [only marks]
                [mark options={solid,mark = +, mark size=3pt, black}]            
                table[x=BcX(T),y=IcX(kA),col sep=comma, dashed] 
                {Data/Ic_Fit.csv};  
                
			\legend{$\SI{10}{\kelvin}$,
			        $\SI{20}{\kelvin}$,
			        $\SI{30}{\kelvin}$,
			        $\SI{40}{\kelvin}$,
			        $\SI{50}{\kelvin}$,
			        $\SI{60}{\kelvin}$,
			        $\SI{70}{\kelvin}$,
			        $\SI{80}{\kelvin}$,
			        $\SCAl{B}{\RM{p}}$,
			        {intersection},
			        }             
		\end{axis}		
	\end{tikzpicture}
}

\newcommand{\FigIcFitFunT}[0]{
    	\begin{tikzpicture}
		\begin{axis}[	
            temflineplot,
            width       = 8cm, 
            height      = 6cm,
            xmin        = 0, 
            xmax        = 90,
            xtick       = {0,10,...,90},
            ymin        = 0, 
            ymax        = 12,
            ytick       = {0,2,...,12},
            xlabel      = $\SCA{T}$, 
            ylabel      = $\SCAl{I}{\RM{c}}$,
            x unit      = \si{\kelvin}, 
            y unit      = \si{\kilo\ampere},       
            ]            
            \addplot+ 
                [only marks]
                [mark options={solid,mark= +, mark size=3pt, black}]
                table[x=T(K),y=IcMeas(kA),col sep=comma] 
                {Data/Ic_Fit.csv}; 
                
            \pgfplotsset{cycle list shift=-1} 
                
            \addplot+ 
                [no markers] 
                table[x=T(K),y=IcFit0(kA),col sep=comma] 
                {Data/Ic_Fit.csv}; 
            \addplot+ 
                [no markers] 
                table[x=T(K),y=IcFit45(kA),col sep=comma] 
                {Data/Ic_Fit.csv}; 
            \addplot+ 
                [no markers] 
                table[x=T(K),y=IcFit75(kA),col sep=comma] 
                {Data/Ic_Fit.csv}; 
                
			\legend{{measurement},
		        	$\RMtheta_\RM{B}=0^{\circ}$,
			        $\RMtheta_\RM{B}=45^{\circ}$,
			        $\RMtheta_\RM{B}=75^{\circ}$,
			        }             
		\end{axis}		
	\end{tikzpicture}
}

\newcommand{\FigIcFitLiftFc}[0]{
    	\begin{tikzpicture}
		\begin{axis}[	
            temflineplot,
            legend style={yshift=-90pt},
            width       = 8cm, 
            height      = 6cm,
            xmin        = 0, 
            xmax        = 90,
            xtick       = {0,10,...,90},
            ymin        = 0, 
            ymax        = 2,
            ytick       = {0,0.25,...,2},
            xlabel      = $\SCA{T}$, 
            ylabel      = $f_{\RM{c}}$,
            x unit      = \si{\kelvin}, 
            y unit      = \si{-},        
            ]            
            
            \addplot+ 
                [no markers] 
                table[x=T(K),y=lf0,col sep=comma] 
                {Data/Ic_Fit.csv}; 
            \addplot+ 
                [no markers] 
                table[x=T(K),y=lf45,col sep=comma] 
                {Data/Ic_Fit.csv}; 
            \addplot+ 
                [no markers] 
                table[x=T(K),y=lf75,col sep=comma] 
                {Data/Ic_Fit.csv}; 
            
			\legend{$\RMtheta_\RM{B}=0^{\circ}$,
			        $\RMtheta_\RM{B}=45^{\circ}$,
			        $\RMtheta_\RM{B}=75^{\circ}$,
			        {meas},
			        }             
		\end{axis}		
	\end{tikzpicture}
}

\newcommand{\FigIronHysteresis}[0]{
    	\begin{tikzpicture}
		\begin{axis}[
		    restrict x to domain=-5:5,
            temflineplot,
            legend pos  = south east,
            width       = 8cm, 
            height      = 6cm,
            xmin        = -2, 
            xmax        = 2,
            ymin        = -2, 
            ymax        = 2,
            xtick    = {-2,-1.5,...,2},
            ytick    = {-2,-1.5,...,2},
            xlabel      = $\SCA{H}$, 
            ylabel      = $\SCA{B}$,
            x unit      = \si{\kilo\ampere\per\meter}, 
            y unit      = \si{\tesla},       
            ]            
            \addplot+ 
                [dashed, black]
                [mark options={solid,mark = diamond*, mark size=1pt, black}] 
                table[x=Hfirst(A/m),y=Bfirst(T),col sep=comma] 
                {Data/IronHysteresis.csv}; 
                            
            \addplot+ 
                [no markers] 
                table[x=Hhyst(A/m),y=Bup(T),col sep=comma] 
                {Data/IronHysteresis.csv};
                
            \pgfplotsset{cycle list shift=-1} 
            
            \addplot+ 
                [no markers] 
                table[x=Hhyst(A/m),y=Bdn(T),col sep=comma] 
                {Data/IronHysteresis.csv};               			
		
			\legend{{first magnetization},
			        {Jiles-Atherton} 
			        }
		\end{axis}
	\end{tikzpicture}	
}

\newcommand{\FigIronHysteresisMultipoles}[0]{
    	\begin{tikzpicture}
		\begin{axis}[
            temflineplot,
            legend pos  = north east,
            width       = 8cm, 
            height      = 6cm,
            xmin        = 0, 
            xmax        = 5,
            ymin        = 0, 
            ymax        = 25,
            xtick    = {0,0.5,...,5},
            ytick    = {0,5,...,25},
            xlabel      = $\SCA{I}$, 
            ylabel      = $|\RMDelta\SCAl{b}{1}|$,
            x unit      = \si{\kilo\ampere}, 
            y unit      = \si{units},       
            ]            
                            
            \addplot+ 
                table [x=I (kA),y=db1,col sep=comma] 
                {Data/IronHysteresis.csv}; 
                \label{plot_one}           
                 					
		\end{axis}
		
			\begin{axis}[	
            temflineplot,
		    axis y line*= right,
		    xmajorgrids = false,
	    	ymajorgrids = false,
            legend pos  = north east,
            width       = 8cm, 
            height      = 6cm,
            xmin        = 0, 
            xmax        = 5,
            ymin        = 0, 
            ymax        = 2,
            ytick    = {0,1,...,2},
            xlabel      = , 
            ylabel      = $|\RMDelta\SCAl{b}{3,5,7}|$,
            x unit      = , 
            y unit      = \si{units}, 
            hide x axis
            ]  
            
            \addlegendimage{/pgfplots/refstyle=plot_one}\addlegendentry{$\RMDelta\SCA{b1}$} 
            
            \pgfplotsset{cycle list shift=+1}
            
            \addplot+ 
                table[x=I (kA),y=db3,col sep=comma] 
                {Data/IronHysteresis.csv};
                \label{plot_two} 
            \addplot+ 
                table[x=I (kA),y=db5,col sep=comma] 
                {Data/IronHysteresis.csv}; 
                \label{plot_three} 
            \addplot+ 
                table[x=I (kA),y=db7,col sep=comma] 
                {Data/IronHysteresis.csv};  
                \label{plot_four} 
                        
            \addlegendimage{/pgfplots/refstyle=plot_two}\addlegendentry {$\RMDelta\SCA{b3}$} 
            \addlegendimage{/pgfplots/refstyle=plot_three}\addlegendentry {$\RMDelta\SCA{b5}$} 
            \addlegendimage{/pgfplots/refstyle=plot_four}\addlegendentry {$\RMDelta\SCA{b7}$} 
            
		\end{axis}

	\end{tikzpicture}	
}

\newcommand{\FigStaircase}[0]{
    \begin{tikzpicture}
		\begin{axis}[	
            temflineplot,	    
            width       = 8cm, 
            height      = 6cm,
            xmin        = -1000, 
            xmax        = 3000,
            ymin        = 0, 
            ymax        = 2,	
            ytick       = {0,0.5,...,2},
            xlabel      = $\SCA{t}$, 
            ylabel      = $\SCA{I}$,
            x unit      = \si{\second}, 
            y unit      = \si{\kilo\ampere},              
            ]                      
            \addplot+ 
                [dashed, no markers] 
                table[x=tPre,y=IPre,col sep=comma, dashed] 
                {Data/Staircase.csv};            
            \addplot+ 
                [no markers] 
                table[x=tTest,y=ITest,col sep=comma, dashed] 
                {Data/Staircase.csv};              
            \addplot+ 
                [ultra thin, only marks, mark options={mark = +, mark size=3pt, black}] 
                table[x=tmeas_up,y=Imeas_up,col sep=comma] 
                {Data/Staircase.csv};   
            \addplot+ 
                [ultra thin, only marks, mark options={mark = x, mark size=3pt, black}] 
                table[x=tmeas_dn,y=Imeas_dn,col sep=comma] 
                {Data/Staircase.csv};      
            \addplot+ 
                [mark=none, dashed, thin, gray!100] 
                coordinates {(0,0) (0,2)}; 
			
			\legend{
			        pre$\HYPHEN$cycle,
			        staircase,
			        up,
			        down,
			        } 
		\end{axis}						
	\end{tikzpicture}
}

\newcommand{\FigJnormDistributionInCable}[0]{
    \begin{tikzpicture}
    
        \node
            [inner sep=0pt] 
            at (0,0)
            {\includegraphics[width=8.0cm]{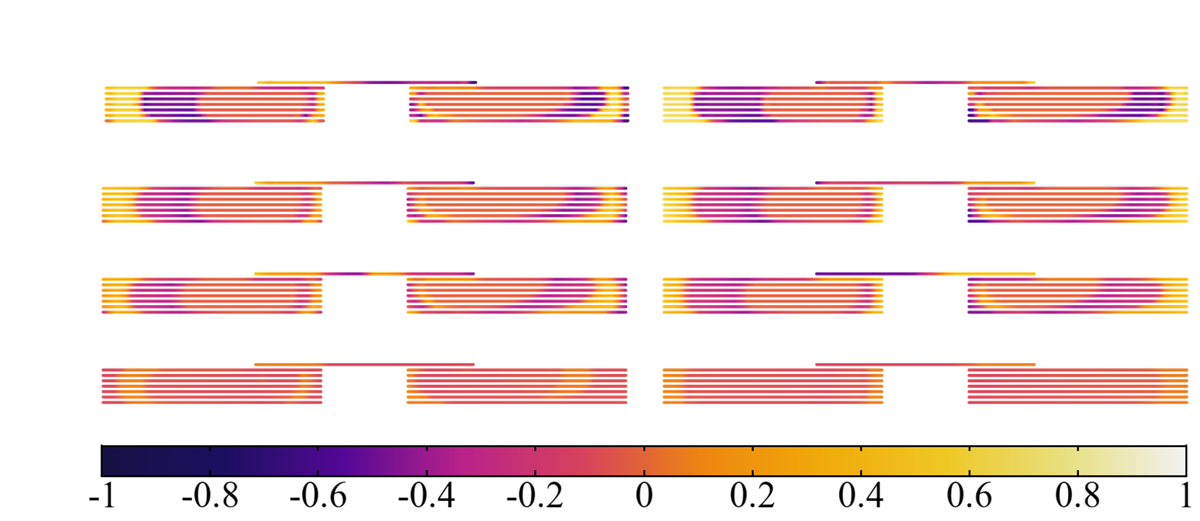}};       
        \node 
            [anchor=center]
            at (-1.625, 1.6) 
            {\scriptsize {ramp$\HYPHEN${up}, $\SI{1}{\kilo\ampere}$}}; 
        \node 
            [anchor=center]
            at (2.3, 1.6) 
            {\scriptsize {ramp$\HYPHEN${down}, $\SI{1}{\kilo\ampere}$}};             
        \node 
            [anchor=east]
            at (-3.5, 1.08) 
            {\scriptsize ${\SI{4.5}{\kelvin}}$}; 
        \node 
            [anchor=east]
            at (-3.5, 0.38) 
            {\scriptsize ${\SI{9}{\kelvin}}$}; 
        \node
            [anchor=east]
            at (-3.5, -0.28) 
            {\scriptsize ${\SI{25}{\kelvin}}$};  
        \node
            [anchor=east]
            at (-3.5, -0.92) 
            {\scriptsize ${\SI{68}{\kelvin}}$}; 

	\end{tikzpicture}
}

%% file: Table/TAB_SourceCode.tex
\newcommand{\TabHTSTapeParameters}[0]{
        \begin{tabular}{lccl}
            \toprule
            Name & Unit & Value & Description \\
            \midrule
            $\RM{\ITdelta_{w}}$ & [$\si{\milli\meter}$] & $\SI{10}{}$ & Tape width \\
            $\RM{\ITdelta_{t}}$ & [$\si{\milli\meter}$] & $\SI{1e-6}{}$ & Tape thickness \\
            $\SCAl{J}{\RM{c}}$  & [$\si{\kilo\ampere\per\milli\meter\squared}$] & $\SI{100}{}$ & Critical current density \\
            \bottomrule
        \end{tabular}    
}

\newcommand{\TabTapeSpecifications}[0]{
        \begin{tabular}{lccl}
            \toprule
            Parameter & Unit & Value & Description \\
            \midrule
            Producer  & & &  {S}unam~\cite{SUNAM2019website} \\
            Technology & & & {IBAD}~\cite{iijima1992plane,reade1992laser} \\
            Substrate  & & & Hastelloy \\
            Stabilizer & & & Copper \\
            $\ITdelta_{\RM{t,sub}}$  & [$\si{\micro\meter}$] & 100 & Substrate thickness\\
            $\ITdelta_{\RM{t,stab}}$ & [$\si{\micro\meter}$] & 40  & Stabilizer thickness\\
            $\ITdelta_{\RM{t}}$ & [$\si{\micro\meter}$] & 150 & Tape thickness\\
            $\ITdelta_{\RM{w}}$ & [$\si{\milli\meter}$] & 5.5  & Tape width\\
            $\RM{I_{c,meas}}$  & [$\si{\ampere}$] & 300 & $@$ $\SI{77}{\kelvin}$, self-field\\
            $\RM{J_{c}}(\VEC{B},\SCA{T})$ & [$\si{\ampere\per\milli\meter\squared}$] & fit & Fit in~\cite{fleiter2014characterization}\\
            ${n}$ & [-] & ${4}\leq{n}\leq{30}$ & Power-law index\\
            \bottomrule
        \end{tabular}           
}
\newcommand{\TabIcFitParameters}[0]{
        \begin{tabular}{lcc}
        \toprule
            Name & Unit & Value \\
        \midrule
            $\RM{g_0}$ & - & $0.03$ \\
            $\RM{g_1}$ & - & $0.25$ \\ 
            $\RM{g_2}$ & - & $0.06$ \\
            $\RM{g_3}$ & - & $0.06$ \\
            $\RM{T_{c0}}$ & $\si{\kelvin}$ & $93$ \\
            $\RM{p_c}$ & - & $0.5$ \\
            $\RM{q_c}$ & - & $2.5$ \\
            $\RM{B_{i0c}}$ & $\si{\tesla}$ & $140$ \\
            ${\ITgamma_\RM{c}}$ & - & $2.44$ \\
            ${\ITalpha_\RM{c}}$ & $\frac{\si{\mega\ampere\tesla}}{\si{\milli\meter^{2}}}$ & $1.86$ \\
        \bottomrule
        \end{tabular}  
        \hspace{0.25cm}
        \begin{tabular}{lcc}
        \toprule
            Name & Unit & Value \\
        \midrule
            $\ITnu$ & - & $1.85$  \\
            $\RM{a}$ & - & $0.1$  \\ 
            $\RM{n_0}$ & - & $1$  \\
            $\RM{n_1}$ & - & $1.4$  \\
            $\RM{n_2}$ & - & $4.45$ \\
            $\RM{p_{ab}}$ & - & $1$ \\
            $\RM{q_{ab}}$ & - & $5$ \\
            $\RM{B_{i0ab}}$ & $\si{\tesla}$ & $250$ \\
            ${\ITgamma_\RM{ab}}$ & - & $1.63$ \\
            ${\ITalpha_\RM{ab}}$ & $\frac{\si{\mega\ampere\tesla}}{\si{\milli\meter^{2}}}$ & $68.3$ \\
        \bottomrule
        \end{tabular}      
}

\newcommand{\TabJilesAthertonParameters}[0]{
        \begin{tabular}{lccl}
            \toprule
            Name & Unit & Value & Description \\
            \midrule
            $\RM{M_{s}}$  & [$\si{\ampere\per\meter}$] & $\SI{1.35e6}{}$ & Saturation magnetization \\
            $\RM{a}$      & [$\si{\ampere\per\meter}$] & 90 & Domain wall density \\
            $\RM{k}$      & [$\si{\ampere\per\meter}$] & 40 & Pinning loss \\
            $\RM{c}$      & [--]                        & $\SI{1e-6}{}$ & Magnetization reversibility\\
            $\RMalpha$    & [--]                        & $\SI{50e-6}{}$  & Inter-domain coupling \\
            \bottomrule
        \end{tabular}    
}

\newcommand{\TabAnalysisScenarios}[0]{
        \begin{tabular}{cccc}
           \toprule
            Scenario & $\SCAl{T}{\RM{op}}$ [\si{\kelvin}] & $\SCAl{I}{\RM{p}}$ [\si{\kilo\ampere}] & $\SCAl{B}{\RM{p}}$ [\si{\tesla}]\\
            \midrule
            1 & 4.5  & 5     & 2.5 \\
            2 & 9    & 4.75  & 2.4 \\
            3 & 25   & 3.75  & 2.0 \\
            4 & 68   & 1.75  & 1.0  \\
            \bottomrule
        \end{tabular}           
}

%% file: Util/UTIL_mathGreekVariablesV0_0.tex
\usepackage{textgreek}

\newcommand{\ITalpha}[0]{\alpha}

\newcommand{\ITgamma}[0]{\gamma}
\newcommand{\ITdelta}[0]{\delta}

\newcommand{\ITmu}[0]{\mu}
\newcommand{\ITnu}[0]{\nu}

\newcommand{\ITpi}[0]{\pi}
\newcommand{\ITrho}[0]{\rho}

\newcommand{\ITsigma}[0]{\sigma}

\newcommand{\ITvarphi}[0]{\varphi}
\newcommand{\ITchi}[0]{\chi}

\newcommand{\RMalpha}[0]{\textrm{\textalpha}}

\newcommand{\RMdelta}[0]{\textrm{\textdelta}}
\newcommand{\RMepsilon}[0]{\textrm{\textepsilon}}

\newcommand{\RMtheta}[0]{\textrm{\texttheta}}

\newcommand{\RMomega}[0]{\textrm{\textomega}}

\newcommand{\RMGamma}[0]{\Gamma}
\newcommand{\RMDelta}[0]{\Delta}

\newcommand{\RMPhi}[0]{\Phi}

\newcommand{\RMOmega}[0]{\Omega}

%% file: Util/UTIL_mathOperatorsV0_0.tex
\newcommand{\HYPHEN} [0] {\operatorname{-}}
\newcommand{\RM}     [1] {\mathrm{#1}}
\newcommand{\BS}     [1] {\boldsymbol{#1}}
\newcommand{\BAR}    [1] {\overline{#1}}

\newcommand{\VEC}   [1] {\protect\vphantom{#1}\smash{\BS{\mathbf{#1}}}}
\newcommand{\VECl}  [2] {\protect\vphantom{#1}\smash{\BS{\mathbf{#1}}}_{#2}}
\newcommand{\VECu}  [2] {\protect\vphantom{#1}\smash{\BS{\mathbf{#1}}}^{#2}}

\newcommand{\SCA}   [1] {\protect\vphantom{#1}\smash{\RM{#1}}}
\newcommand{\SCAl}  [2] {\protect\vphantom{#1}\smash{\RM{#1}}_{#2}}
\newcommand{\SCAu}  [2] {\protect\vphantom{#1}\smash{\RM{#1}}^{#2}}
\newcommand{\SCAlu} [3] {\protect\vphantom{#1}\smash{\RM{#1}}_{#2}^{#3}}

\newcommand{\GRAD} [0] {\nabla}

\newcommand{\CURL} [0] {\nabla\times}

\newcommand{\D}     [1] {\!\mathop{}\mathrm{d}{#1}}
\newcommand{\dxP}   [1] {\partial_{#1}}


%% file: Util/UTIL_mathEquations.tex
\newcommand{\EQdefJpowerLaw}[0]
    { 
    \SCA{\ITrho}(|\VEC{J}|) =
    \frac{\SCAl{E}{\RM{c}}}{\SCAl{J}{\RM{c}}}
    \left(\frac{|\VEC{J}|}{\SCAl{J}{\RM{c}}}\right)^{{n}-1}
    }
       
\newcommand{\EQstrongAampereMaxwell}[0]
    { 
    \CURL\ITmu^{-1}\CURL\VECu{A}{\star} 
    + \SCA{\ITsigma}\dxP{t}\VECu{A}{\star} 
    & = 0
    }
   
\newcommand{\EQstrongHfaraday}[0]
    { 
    \CURL\SCA{\ITrho}\CURL\VEC{H} 
    + \dxP{t}\SCA{\ITmu}\VEC{H} 
    - \CURL\VEC{\ITchi}\SCAl{u}{\RM{s}} 
    & = 0
    }
   
\newcommand{\EQconstraintIsource}[0]
    { 
    \int\limits_{\SCAl{\RMOmega}{\RM{H}}}\!\!
    \VEC{\ITchi}\cdot(\CURL\VEC{H})
    \D{\SCA{\RMOmega}} 
    }

\newcommand{\EQdefStrongLaplaceAz}[0]
    {
    \SCAl{A}{\RM{z}}(r,\ITvarphi)= 
    \sum_{k=1}^{\infty}{r}^{k}
    (\SCAl{\cal{A}}{k} \sin{k}\ITvarphi + \SCAl{\cal{B}}{k} \cos{k}\ITvarphi)
    }
    
\newcommand{\EQdefStrongLaplaceBr}[0]
    {
    \SCAl{B}{r}(r,\ITvarphi) &= 
    \sum_{k=1}^{\infty}{k}{r}^{k-1} 
    (\SCAl{\cal{A}}{k} \cos{k}\ITvarphi - \SCAl{\cal{B}}{k} \sin{k}\ITvarphi)
    }
    
\newcommand{\EQdefStrongLaplaceBphi}[0]
    {
    \SCAl{B}{\ITvarphi}(r,\ITvarphi) &= 
    - \sum_{k=1}^{\infty}{k}{r}^{k-1} 
    (\SCAl{\cal{A}}{k} \sin{k}\ITvarphi + \SCAl{\cal{B}}{k} \cos{k}\ITvarphi)
    }
    
\newcommand{\EQdefFourierCoeffAk}[0]
    {
    \SCAl{A}{k}(\SCAl{r}{0}) &= 
    \frac{1}{\pi} \int_{0}^{2\pi}
    \SCAl{B}{r}(\SCAl{r}{0},\ITvarphi) 
    \cos{k}\ITvarphi\ {d}\ITvarphi
    }

\newcommand{\EQdefFourierCoeffBk}[0]
    {
    \SCAl{B}{k}(\SCAl{r}{0}) &= 
    \frac{1}{\pi} \int_{0}^{2\pi} 
    \SCAl{B}{r}(\SCAl{r}{0},\ITvarphi) 
    \sin{k}\ITvarphi \ {d}\ITvarphi
    }
    
\newcommand{\EQdefFourierCoeffck}[0]
    {
    \SCAl{c}{k}(\SCAl{r}{0}) = 
    \SCAl{b}{k}(\SCAl{r}{0}) + i \SCAl{a}{k}(\SCAl{r}{0}) 
    = 10^4 \frac{\SCAl{C}{k}(\SCAl{r}{0})}{\SCAl{B}{\RM{K}}}
    }
    
\newcommand{\EQdefFourierTHD}[0]
    {
    \SCAl{F}{\RM{THD}}(\SCAl{r}{0}) = 
    \sqrt{
    \sum_{{k}=1;\ {k}\neq \SCA{K}}^{\infty} 
    \SCAlu{b}{k}{2}(\SCAl{r}{0}) + \SCAlu{a}{k}{2}(\SCAl{r}{0})}
    }
    
\newcommand{\EQdefTapeExtFieldLosses}[0]
    {
    \SCAl{w}{\RM{J}} =
    \SCAl{\RMdelta}{\RM{w}}
    \SCAl{J}{\RM{c}}\SCAl{B}{\RM{c}}
    \left(\frac{2}{\SCAl{b}{\RM{p}}}
    \RM{ln}(\RM{cosh}\ \SCAl{b}{\RM{p}})
     -\RM{tanh}\ \SCAl{b}{\RM{p}}\right)
    }
    
\newcommand{\EQdefTapeSelfFieldLosses}[0]
    {
    \SCAl{w}{\RM{J}} =           
    \frac{\SCAl{\ITmu}{0}}{\SCAl{\RMdelta}{\RM{w}}\SCAl{\RMdelta}{\RM{h}}} 
    \SCAlu{I}{\RM{c}}{2}
    \frac{\SCAlu{i}{\RM{p}}{4}}{{6}\ITpi}
    }
    
\newcommand{\EQdefTHDerrorMesh}[0]
    {
    \SCAl{\RMepsilon}{\SCA{\RMDelta{x}}}
    = \frac{|\SCAl{F}{\RM{THD}}-\SCAlu{F}{\RM{THD}}{\SCA{\RMDelta{x}}}|}
    {\SCAl{F}{\RM{THD}}}
    }

\newcommand{\EQdefCorrFactorJc}[0]
    {
    \SCAl{f}{\RM{c}}(\SCA{T}) = 
    \left.\frac{\SCAl{I}{\RM{c,meas}}(\SCA{T})}
    {\SCAl{J}{\RM{c}}
    (\SCAl{B}{\RM{p,coil}}(\SCA{T}),\SCA{T},\SCAl{\RMtheta}{\RM{B}})
   \SCAl{S}{\RM{HTS}}}
    \right\rvert_{\SCAl{\RMtheta}{\RM{B}}=0^{\circ}}
    }
    
\newcommand{\EQdefRotCoilPhi}[0]
    {
    \RMPhi(t) &= 
    \sum_{k=1}^{\infty}  
    \SCAl{f}{\RM{s}}
    \left[\SCAl{A}{k}(\SCAl{r}{\RM{c0}}) \cos k\ITvarphi
    -\SCAl{B}{k}(\SCAl{r}{\RM{c0}}) \sin k\ITvarphi \right]
    }
    
\newcommand{\EQdefRotCoilSensFactor}[0]
    {
    \SCAl{f}{\RM{s}}(k) &=
    \frac{{2} \SCAl{N}{\RM{c}} \SCAl{l}{\RM{c}} \SCAl{r}{\RM{c0}}}
    {k}
    }

%% file: Section/SEC01_Abstract.tex

\begin{abstract}
Screening currents are field-induced dynamic phenomena which occur in superconducting materials, leading to persistent magnetization. Such currents are of importance in \mbox{{R}e{BCO}} tapes, where the large size of the superconducting filaments gives rise to strong magnetization phenomena. In consequence, superconducting accelerator magnets based on \mbox{{R}e{BCO}} tapes might experience a relevant degradation of the magnetic field quality in the magnet aperture, eventually leading to particle beam instabilities. Thus, persistent magnetization phenomena need to be accurately evaluated. In this paper, the 2D finite element model of the \FEATHER{} magnet is presented. The model is used to analyze the influence of the screening current-induced magnetic field on the field quality in the magnet aperture. The model relies on a coupled field formulation for eddy current problems in time-domain. The formulation is introduced and verified against theoretical references. Then, the numerical model of the \FEATHER{} magnet is detailed, highlighting the key assumptions and simplifications. The numerical results are discussed and validated with available magnetic measurements. A satisfactory agreement is found, showing the capability of the numerical tool in providing accurate analysis of the dynamic behavior of the \FEATHER{} magnet.
\\
\\
\textbf{Index Terms --} High-temperature superconductors, screening currents, magnetic fields, magnetization, finite-element analysis, superconducting coils, accelerator magnets.
           
\end{abstract}



    

%% file: Section/SEC02_Body.tex

\section{Introduction} 
    \label{SEC_Introduction}
High Temperature Superconducting (HTS) materials are a promising technology for high-field magnets in particle accelerators. In particular, superconducting tapes based on {R}e{BCO} compounds~\cite{wu1987superconductivity} have a critical temperature of $\SI{93}{\kelvin}$ and an estimated upper critical field of $\SI{140}{\tesla}$~\cite{golovashkin1991low}. These properties are about one order of magnitude higher than in traditional Low Temperature Superconducting (LTS) materials, such as $\RM{Nb{\HYPHEN}Ti}$ or $\RM{Nb_{3}Sn}$~\cite{wilson1983superconducting}. Thus, HTS materials might be used in building accelerator magnets with higher magnetic fields and thermal margins~\cite{van2016high}. A significant milestone in this direction was recently achieved by the {E}u{CARD}-2~\cite{rossi2015eucard} and {ARIES}~\cite{ARIES2019website} projects, which led to the construction of the HTS accelerator dipole insert-magnet \FEATHER{}~\cite{kirby2014accelerator,van2014study}. This demonstrator magnet is designed to operate inside the aperture of the $\RM{Nb_{3}Sn}$ {FRESCA2} dipole magnet~\cite{milanese2011design,ferracin2013development,rochepault2017mechanical,willering2018cold,willering2019tests},
producing a peak field of $\SI{5}{\tesla}$ at a nominal current of $\SI{10}{\kilo\ampere}$, in a background field of $\SI{13}{\tesla}$. The magneto-thermal behavior of the \FEATHER{} magnet was recently tested in a stand-alone configuration~\cite{van2018powering}, and the influence of the superconducting coil dynamics on the magnetic transfer function was measured~\cite{petrone2018measurement}.

One of the key requirements for accelerator magnets is to produce high-quality magnetic fields in their magnet aperture (see e.g.~\cite{bruning2004lhc}), as field imperfections can lead to particle beam instabilities~\cite{shi2000collective}. Therefore, the current density distribution within the superconductor should be as uniform as possible. Conversely, HTS tapes behave as wide and anisotropic mono-filaments, resulting in the dynamic regime in screening currents which are persistent, as they flow in a superconducting material. These currents prevent a homogeneous current density distribution by producing a persistent magnetization in the tape, potentially degrading the magnetic field quality~\cite{uglietti2010measurements,yanagisawa2010magnitude,amemiya2015magnetisation,dilasser2016experimental,wang2016screening,fazilleau2018screening,noguchi2019simple}. Attempts in reducing persistent magnetization phenomena either by tape striation~\cite{nast2014influence} or tape-field alignment~\cite{van2016measurement} were not yet fully satisfactory.

The screening currents magnitude is determined by the operational margin of the tape, i.e. the difference between the supply and the critical current, which limits the superconducting state. As a consequence, persistent magnetization phenomena are more severe at low current. This poses a major challenge for high-field accelerator magnets, whose supply current typically varies over one order of magnitude during the energy ramp. For this reason, persistent magnetization phenomena need to be carefully evaluated, and possibly predicted by means of numerical models, as they may limit the use of HTS technology in accelerator magnets.

In this paper, we present the time domain analysis of the magnetic field quality in the \FEATHER{} magnet. A dedicated 2D numerical model is developed using the finite element method (FEM, e.g.~\cite{zienkiewicz2005finite}). The model implements a coupled $\VEC{A}$-$\VEC{H}$ field formulation~\cite{dular1997magnetostatic,biro1999edge} for HTS materials~\cite{brambilla2018finite,dular2019finite}, extended to the simulation of HTS magnets~\cite{bortot2020coupled}. This is achieved by following a domain decomposition strategy, solving the field problem for the magnetic field strength $\VEC{H}$ in the superconducting regions, and for the magnetic vector potential $\VEC{A}$ in the normal-conducting and non-conducting regions. Thus, the coupled field formulation accounts for electrodynamic phenomena in the superconducting coil by solving an eddy current problem in time-domain. The advantages of this approach are discussed in~\cite{dular2019finite}.

The model of the \FEATHER{} magnet is used to quantify the contribution of the screening currents-induced magnetic field to the magnetic field quality. Moreover, simulations provide the current density distribution within each superconducting tape, which is crucial for the determination and understanding of the Joule losses and the Lorentz forces in the coil. The numerical results are compared with measurements of the magnetic field quality in the aperture of the \FEATHER{} magnet. In this comparison, a high degree of consistency is found.

The paper is organized as follows. The mathematical model is discussed in Section~\ref{SEC_MathematicalModel} and verified in Section~\ref{SEC_VerificationOfTheMathematicalModel} by comparing simulations of single tapes with theoretical references. In Section~\ref{SEC_FeatherModel}, the numerical model of the \FEATHER{} magnet is presented, highlighting assumptions and simplifications. The validation of the model with measurements is reported in Section~\ref{SEC_SimWithMeas}, discussed in Section~\ref{SEC_Discussion} and followed by the conclusions.

\section{Mathematical Model}
    \label{SEC_MathematicalModel}
HTS materials exhibit a highly nonlinear electric field-current density relation~(e.g.~\cite{wu1987superconductivity}), which determines the magnetic field penetration in the superconductors and, ultimately, the dynamics of the screening currents. Thus, the mathematical model of such relation is crucial for the time-domain analysis of the \FEATHER{} magnet.

The resistivity $\ITrho$ in HTS materials can be modeled by means of a phenomenological percolation-depinning law proposed in~\cite{yamafuji1997current}. The law introduces a lower limit for the current density, below which the magnetic field is frozen in the superconductor and flux creep~\cite{wilson1983superconducting} cannot occur. However, the current density values used in practical applications are typically much higher than the lower limit considered in the percolation law. Thus, a further simplification into a power law~\cite{rhyner1993magnetic} is sufficient, as shown in~\cite{sirois2018comparison}, and is adopted in this work. The resistivity reads
    \begin{align}
        \label{EQ_powerLaw}
            \EQdefJpowerLaw{},      
    \end{align}
where $\VEC{J}$ is the current density, $\SCAl{E}{\RM{c}}$ is the critical electric field strength, set to $\SI{1e-4}{\volt\per\meter}$~\cite{dew1988model}, and the material- and field-dependent parameters $\SCAl{J}{\RM{c}}$ and ${n}$ are the critical current density and the power-law index, respectively. For the limiting cases of $n\to0$ and $n\to\infty$, the power law approximates the behavior of normal conducting materials and superconducting materials in critical state~\cite{bean1962magnetization,bean1964magnetization}.

At low currents, i.e. $|\VEC{J}|\to0$, the resistivity in~\eqref{EQ_powerLaw} vanishes. Thus, the field problem is formulated avoiding the electrical conductivity $\ITsigma$ in superconducting domains~\cite{ruiz2004numerical,dular2019finite}, and the electrical resistivity in non-conducting domains, such that the material properties remain finite. This is done by combining a domain decomposition strategy with a dedicated coupled field formulation (see~\cite{bortot2020coupled}), briefly discussed in the next section.


\subsection{Formulation of the Field Problem}
    \label{SEC_FormulationOfTheFieldProblem}
%
\begin{figure}[tb]
  \centering
	\includegraphics[width=8.0cm]{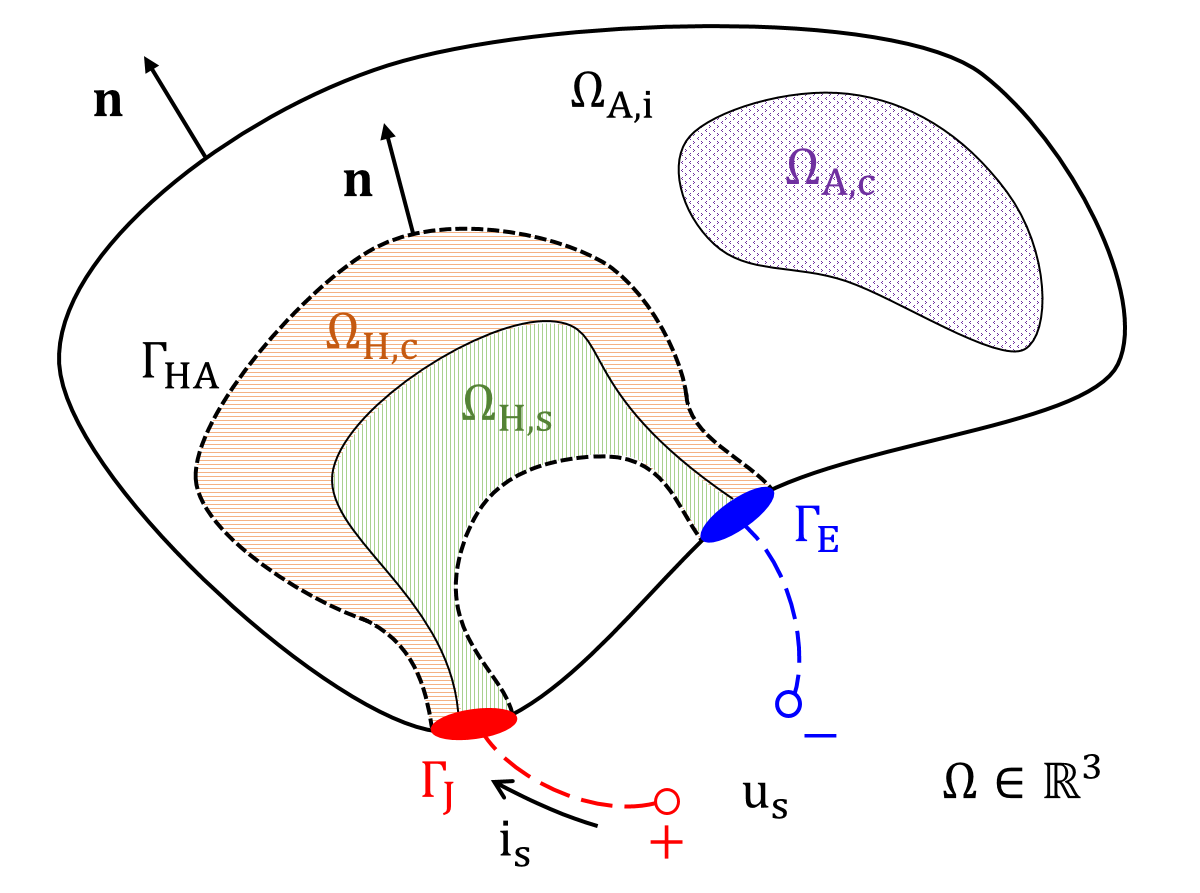}
	\caption{Decomposition of the domain $\SCA{\RMOmega}$ into the source and source-free domains $\SCAl{\RMOmega}{\RM{H}}$~$=\SCAl{\RMOmega}{\RM{H,s}}\!\cup\SCAl{\RMOmega}{\RM{H,c}}$ and $\SCAl{\RMOmega}{\RM{A}}$~$=\SCAl{\RMOmega}{\RM{A,c}}\!\cup\SCAl{\RMOmega}{\RM{A,i}}$. The two domains are separated by the interface $\SCAl{\RMGamma}{\RM{HA}}$, shown as a dashed line. The regions $\SCAl{\RMOmega}{\RM{H,s}}$ and $\SCAl{\RMOmega}{\RM{H,c}}$ represent the superconducting and normal-conducting parts in the source domain. The regions $\SCAl{\RMOmega}{\RM{A,c}}$ and $\SCAl{\RMOmega}{\RM{A,i}}$ show the normal-conducting and non-conducting parts in the source-free domain. The electrical ports are marked as $\SCAl{\RMGamma}{\RM{J}}$ and $\SCAl{\RMGamma}{\RM{E}}$. This figure is based on Fig.~2 from~\cite{bortot2020coupled}.}
	\label{FIG_DomainDecomposition}
\end{figure}
The computational domain $\SCA{\RMOmega}$ representing the superconducting magnet is illustrated in Fig.~\ref{FIG_DomainDecomposition}. The domain is decomposed into the source and source-free regions $\SCAl{\RMOmega}{\RM{H}}$ and $\SCAl{\RMOmega}{\RM{A}}$ oriented with the unit vector $\VEC{n}$, such that $\SCAl{\BAR{\RMOmega}}{\RM{H}}\cup\SCAl{\BAR{\RMOmega}}{\RM{A}}=\SCA{\BAR{\RMOmega}}$. The source region $\SCAl{\RMOmega}{\RM{H}}$ corresponding to the coil is given by the union of the superconducting and normal-conducting parts $\SCAl{\RMOmega}{\RM{H,s}}$ and $\SCAl{\RMOmega}{\RM{H,c}}$. The source-free region $\SCAl{\RMOmega}{\RM{A}}$, containing the remainder of the magnet such as the iron yoke, the mechanical supports, and the air region, is given by the union of the normal-conducting and non-conducting parts $\SCAl{\RMOmega}{\RM{A,c}}$ and $\SCAl{\RMOmega}{\RM{A,i}}$. A constant magnetic permeability $\ITmu$ is assumed in $\SCAl{\RMOmega}{\RM{H}}$, whereas a nonlinear dependency from the magnetic field $\VEC{B}$ is considered for the iron yoke in $\SCAl{\RMOmega}{\RM{A}}$, as $\ITmu(\VEC{B})$. 

The field problem is solved under magnetoquasistatic assumptions for the reduced magnetic vector potential $\VECu{A}{\star}$~\cite{emson1983optimal} in $\SCAl{\RMOmega}{\RM{A}}$, and for the magnetic field strength $\VEC{H}$~\cite{bossavit1988rationale,brambilla2006development} in $\SCAl{\RMOmega}{\RM{H}}$, with suitable boundary conditions on the exterior boundary. The formulation reads
    \begin{align}
        \label{EQstrongAampereMaxwell}
            \EQstrongAampereMaxwell{}\ \text{in}\ \SCAl{\RMOmega}{\RM{A}}, \\
        \label{EQstrongHfaraday}
            \EQstrongHfaraday{}\ \text{in}\ \SCAl{\RMOmega}{\RM{H}}, \\ 
        \label{EQconstraintIsource}
            \EQconstraintIsource{} &= \SCAl{i}{\RM{s}}, 
    \end{align}
where $\VEC{\ITchi}$ is a voltage distribution function~\cite{rodriguez2008voltage,schops2013winding}, $\SCAl{u}{\RM{s}}$ is the source voltage treated as an algebraic unknown, and $\SCAl{i}{\RM{s}}$ is the source current which is imposed via a constraint equation (i.e., a Lagrange multiplier). The sources are provided by means of the electrical ports $\SCAl{\RMGamma}{\RM{J}}$ and $\SCAl{\RMGamma}{\RM{E}}$. The fields $\VECu{A}{\star}$ and $\VEC{H}$ are linked via continuity conditions at the interface of the domains $\SCAl{\RMGamma}{\RM{HA}}$, represented in Fig.~\ref{FIG_DomainDecomposition} by a dashed line. In particular, the continuity of the normal component of the magnetic field $\VECl{B}{\RM{n}}$ and the current density $\VECl{J}{\RM{n}}$, and the tangential component of the magnetic field strength $\VECl{H}{\RM{t}}$ and electric field strength $\VECl{E}{\RM{t}}$ are imposed, ensuring the consistency of the overall field solution.

\subsection{Magnetic Field Quality}
    \label{SEC_MagneticFieldQuality}
The magnetic field quality in accelerator magnets is defined as the set of Fourier coefficients, known also as field harmonics or multipole coefficients. The coefficients are derived from the solution of the field problem in the source-free magnet aperture, which is given by the Laplace equation $\GRAD^{2}\VEC{A}=0$. In the two-dimensional approximation of accelerator magnets, the axial field variations are neglected along the $z$-direction (the longitudinal axis of the magnet). Thus, the field can be expressed as (e.g.~\cite{russenschuck2011field})
    \begin{equation}
        \label{EQ_defStrongLaplaceAz} 
            \EQdefStrongLaplaceAz{}. 
    \end{equation}
where $\SCAl{A}{\RM{z}}$ is the longitudinal component of the magnetic vector potential, $\SCAl{\cal{A}}{k}$ and $\SCAl{\cal{B}}{k}$ are the multipole coefficients, and $(r,\ITvarphi,z)$ are spatial coordinates in a cylindrical reference system consistent with the magnet aperture. The field components are obtained from~\eqref{EQ_defStrongLaplaceAz} as
    \begin{align}
        \label{EQ_defStrongLaplaceBr} 
            \EQdefStrongLaplaceBr{}, \\
        \label{EQ_defStrongLaplaceBphi} 
            \EQdefStrongLaplaceBphi{}.
    \end{align}

The index $k$ represents solutions of the Laplace equation which can be associated to field distributions generated by ideal magnet geometries. As an example, $k$ = 1,2,3 correspond to the dipole, quadrupole, and sextupole field distributions. Once the radial field component~\eqref{EQ_defStrongLaplaceBr} is known at a reference radius $r = \SCAl{r}{0}$ (either by measurements or simulations), the skew and normal multipole coefficients $\SCAl{A}{k}$ and $\SCAl{B}{k}$ are obtained for $k=1,2,3,\ldots$, as  
    \begin{align}
        \label{EQ_defFourierCoeffAk} 
            \EQdefFourierCoeffAk{}, \\
        \label{EQ_defFourierCoeffBk} 
            \EQdefFourierCoeffBk{},
    \end{align}
where the radius $\SCAl{r}{0}$ is usually chosen as 2/3 of the magnet aperture. 

The coefficients are often combined in the complex notation $\SCAl{C}{k}(\SCAl{r}{0})=\SCAl{B}{k}(\SCAl{r}{0})+i\SCAl{A}{k}(\SCAl{r}{0})$ as the skew and normal pairs are orthogonal to each other. Moreover, the coefficients are typically normalized with respect to the main field component $\SCAl{B}{\RM{K}} (\SCAl{r}{0})$, and denoted as $\SCAl{b}{k}$ and $\SCAl{a}{k}$. Thus, the field quality is quantified as a relative error $\SCAl{c}{k}$, for $k=1,2,3,\ldots$, as
    \begin{equation}
        \label{EQ_defFourierCoeffck}  
            \EQdefFourierCoeffck{},
    \end{equation} 
and given in $\SI{1e-4}{}$ units with respect to $\SCAl{B}{\RM{K}}$ at the reference radius $\SCAl{r}{0}$. It is worth mentioning that for reaching accelerator quality standards, the field multipoles shall be limited within a few units~\cite{russenschuck2011field}.

The magnetic field quality can also be conveniently given in terms of total harmonic distortion factor $\SCAl{F}{\RM{THD}}(\SCAl{r}{0})$, which is a scalar quantity defined as
    \begin{equation}
        \label{EQ_defFourierTHD}  
            \EQdefFourierTHD{},
    \end{equation}
where $\SCA{K}$ refers to the index of the main field component.


\section{Verification of the Mathematical Model}
    \label{SEC_VerificationOfTheMathematicalModel}
%
The FEM implementation of the formulation proposed in~\eqref{EQstrongAampereMaxwell}--\eqref{EQconstraintIsource} is used to simulate the dynamic behavior of a single HTS tape, considered as an infinitely thin shell~\cite{krahenbuhl1993thin}. For this simplified case, analytical solutions from previous literature are used for the verification of the numerical results in Sections~\ref{SUBSUBSEC_HysteresisLossExtB} and~\ref{SUBSUBSEC_HysteresisLossSelfB}. Subsequently, a mesh sensitivity analysis is carried out for a known field solution to assess the precision of the code in calculating the multipole coefficients. The mesh sensitivity results are presented in Section~\ref{SUBSEC_MeshSensitivity}.

\subsection{Single Tape Model}
    \label{SUBSEC_SingleTapeModel}
\begin{figure}[tb]
  \centering
	\includegraphics[width=8.0cm]{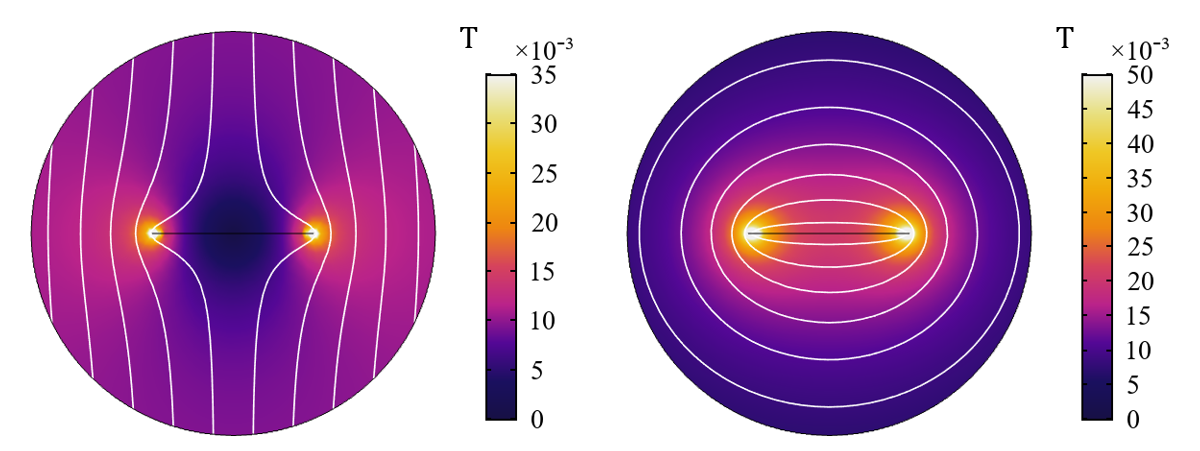}
	\caption{Magnetic field, in $\si{\tesla}$, for a single HTS tape with an $n$-value of 20, at $\SI{1.25}{\sec}$. a) Tape in external sinusoidal field of $\SI{10}{\milli\tesla}$ and frequency of $\SI{1}{\hertz}$. b) Tape in self-field, driven with a sinusoidal current of $\SI{500}{\ampere}$ and frequency of $\SI{1}{\hertz}$.}
	\label{FIG_fieldTape}
\end{figure}

The 2D magnetoquasistatic model of the HTS tape is used for calculating the specific Joule losses per cycle, in the sinusoidal regime. The tape is composed only by one superconducting layer whose specifications are given in Table~\ref{TAB_HTSTapeParameters}. Two scenarios are considered, differing only in the source quantity applied to the tape: 1) An external magnetic field at zero current, (Fig.~\ref{FIG_fieldTape}, left), and 2) a supply current, in self field (Fig.~\ref{FIG_fieldTape}, right). The results are verified against analytical solutions and presented in Sections~\ref{SUBSUBSEC_HysteresisLossExtB} and~\ref{SUBSUBSEC_HysteresisLossSelfB}.

The numerical model of the HTS tape is used over a frequency range of several orders of magnitude. For this reason, an adaptive mesh distribution is used in the tape. The mesh elements are denser at the tape edges, following a geometrical distribution of ratio 25. This allows for resolving the highly-nonlinear current density distribution in the tape. About 500 elements are used for the simulations at low field and current, whereas about 20 elements are used for saturated tapes, in accordance with the relaxation of the magnetic field within the tape. The maximum time step size is given as $\SCAl{\RMDelta{t}}{\RM{max}}=(50{f})^{-1}$, where $f$ is the frequency of the source quantity in the model.
\begin{table}[tb]
    \caption{Tape specifications}
    \label{TAB_HTSTapeParameters}
    \centering
    \TabHTSTapeParameters
\end{table}

\subsubsection{Tape in Perpendicular External Field}
    \label{SUBSUBSEC_HysteresisLossExtB}
A single HTS tape with no supply current is exposed to a time-dependent, perpendicular external field. The layout of the simulated scenario is shown in the box of Fig.~\ref{FIG_AClossQBparamN}. The source term is given by a sinusoidal magnetic field $\SCA{B}(t)=\SCAl{B}{\RM{p}}\RM{sin}(2\pi ft)$, applied perpendicularly to the tape. The specific loss per cycle $\SCAl{w}{\RM{J}}$ is calculated for ${n}=5$, 20 and 40. The case of ${n}=\infty$, which corresponds to the critical state model~\cite{bean1962magnetization,bean1964magnetization}, is calculated analytically. The theory of infinitely thin films with finite width and one dimensional current distribution in a perpendicular field~\cite{brandt1993type,brandt1996superconductors,grilli2013computation} gives $\SCAl{w}{\RM{J}}$ as
    \begin{equation}
        \label{EQ_defTapeExtFieldLosses}
        \EQdefTapeExtFieldLosses{},   
    \end{equation}
where $\SCAl{B}{\RM{c}}={{\ITmu}_{0}}(\SCAl{\RMdelta}{\RM{h}}\SCAl{J}{\RM{c}})/\ITpi$ is the critical magnetic field, $\SCAl{\RMdelta}{\RM{w}}$ and $\SCAl{\RMdelta}{\RM{h}}$ are the width and the thickness of the tape, and $\SCAl{b}{\RM{p}}=\SCAl{B}{\RM{p}}/\SCAl{B}{\RM{c}}$ is the normalized magnetic field.

The losses $\SCAl{w}{\RM{J}}$ are given in Figs.~\ref{FIG_AClossQBparamN} and~\ref{FIG_AClossQfparamN} as a function of $\SCAl{B}{\RM{p}}$ and $f$. The losses converge to the theoretical solution in~\cite{brandt1993type} for increasing ${n}$-values, which is to be expected given that the critical state model corresponds to the power-law equation~\eqref{EQ_powerLaw} where ${n}$ is set to infinite. For low field values, the losses follow a quartic scaling law with respect to the magnetic field amplitude. Once the magnetic field reaches $\SCAl{B}{\RM{c}}$, it fully penetrates in the tape, and the screening current distribution is maintained. The losses grow proportionally with the amplitude of the applied magnetic field, as the model considers the critical current density to be constant and field-independent. For the sake of completeness, Fig.~\ref{FIG_AClossQBparamN} reports also a trend line for a cubic scaling law, which is found in models accounting for finite ${n}$-values and two dimensional current density distributions in the tape~\cite{rabbers2003ac,wuis2009ac,van2016high}. 

The losses presented in Fig.~\ref{FIG_AClossQfparamN} are calculated for a peak field of $\SI{10}{\milli\tesla}$. The field is chosen below the penetration limit, such that the field-screening behavior of the tape is included in the simulation. For high ${n}$-values (see Fig.~\ref{FIG_AClossQfparamN}) the frequency dependency tends to vanish, in accordance with a hysteresis-like behavior, and $\SCAl{w}{\RM{J}}$ converges to the theoretical solution in~\cite{brandt1993type} for ${n}=\infty$. 
\FigureVerificationPlots{}

\subsubsection{Tape in Self-Field}
    \label{SUBSUBSEC_HysteresisLossSelfB}
A source current is imposed to a single HTS tape, in self-field. The layout of the simulated scenario is shown in the box of Fig.~\ref{FIG_AClossQIparamN}. The source term is given by a sinusoidal current $\SCA{I}(t)=\SCAl{I}{\RM{p}}\RM{sin}(2\pi ft)$, applied to the tape as source. The calculation of $\SCAl{w}{\RM{J}}$ is done for ${n}=5$, 20 and 40, whereas the case of ${n}=\infty$ is calculated analytically. The theory of infinitely thin films with finite width and one dimensional current distribution in self-field~\cite{grilli2013computation,norris1970calculation} gives $\SCAl{w}{\RM{J}}$ as
    \begin{equation}
        \label{EQ_defTapeSelfFieldLosses}
        \EQdefTapeSelfFieldLosses{},
    \end{equation}
where $\SCAl{I}{\RM{c}}$ is the critical current of the tape and $\SCAl{i}{\RM{p}}=\SCAl{I}{\RM{p}}/\SCAl{I}{\RM{c}}$ is the normalized supply current. The losses $\SCAl{w}{\RM{J}}$ are given in Figs.~\ref{FIG_AClossQIparamN} and~\ref{FIG_AClossQfparamI} as a function of $\SCAl{I}{\RM{p}}$ and $f$. Consistent with~\eqref{EQ_defTapeSelfFieldLosses}, for high n-values the numerical results show a quartic dependence for currents up to the critical current. Beyond this value, the current density distributes homogeneously in the tape, and the losses are proportional to $\SCAu{I}{{n}+1}$, in accordance with the power-law behavior in~\eqref{EQ_powerLaw}. 

The losses presented in Fig.~\ref{FIG_AClossQfparamI} are calculated for a sub-critical current $\SCAl{I}{\RM{p}}=0.5\ \SCAl{I}{\RM{c}}$. From Fig.~\ref{FIG_AClossQIparamN} it is clear that with increasing ${n}$-value, the simulation results converge to the analytical dependence given in literature~\cite{grilli2013computation,norris1970calculation}, whereas Fig.~\ref{FIG_AClossQfparamI} shows that with increasing ${n}$-value, the frequency dependency vanishes as expected.

\subsection{Mesh Sensitivity}
    \label{SUBSEC_MeshSensitivity}
\begin{figure}[tb]
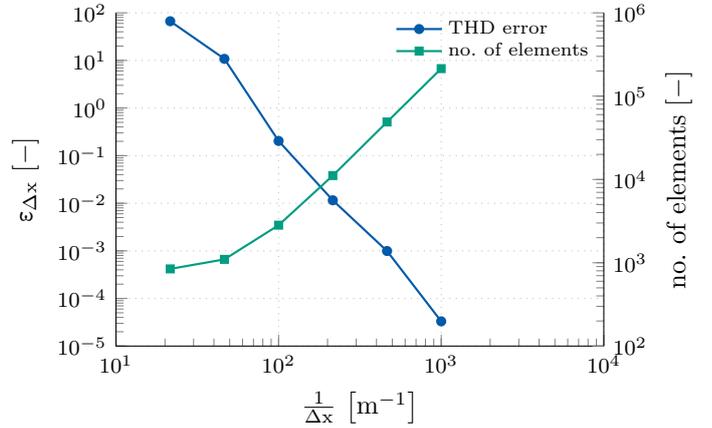

    \centering
    \FigMeshSensitivity         
    \caption{Left axis: relative error in the calculation of the total harmonic distortion index, as function of the reciprocal of the maximum mesh size. Right axis: number of elements in the mesh.}
    \label{FIG_MeshSensitivity}
\end{figure}

In numerical models, the multipole coefficients are obtained by applying the Fast Fourier Transform algorithm to the radial component of the magnetic field, calculated along the reference circumference in the magnet aperture (see Section~\ref{SEC_MagneticFieldQuality}). Care has to be taken, as the finite resolution of the mesh in the spatial discretization introduces a numerical error which affects the calculation of the multipole coefficients~\cite{romer2017defect}. For this reason, a mesh sensitivity analysis is carried out for a reference model where a known analytical field solution is simulated and calculated at the reference circumference. The relative error $\SCAl{\RMepsilon}{\SCA{\RMDelta{x}}}$ is defined as
    \begin{align}
        \label{EQ_defTHDerrorMesh}
            \EQdefTHDerrorMesh{},
    \end{align}
where $\SCAl{F}{\RM{THD}}$ and $\SCAlu{F}{\RM{THD}}{\SCA{\RMDelta{x}}}$ are the total harmonic distortion factors in~\eqref{EQ_defFourierTHD} for the analytical and calculated field solutions. The error is shown in Fig.~\ref{FIG_MeshSensitivity} as a function of the reciprocal of the maximum element size $\SCAl{\RMDelta{x}}{\RM{max}}$. Based on this investigation, triangular elements with $\SCAl{\RMDelta{x}}{\RM{max}}=\SI{1}{mm}$ are chosen for the mesh, yielding an estimated error of $\SI{3e-5}{}$.

\section{Numerical Model of the \FEATHER{} Magnet}
    \label{SEC_FeatherModel}
%
\begin{figure}[tb]
  \centering
	\includegraphics[width=8.0cm]{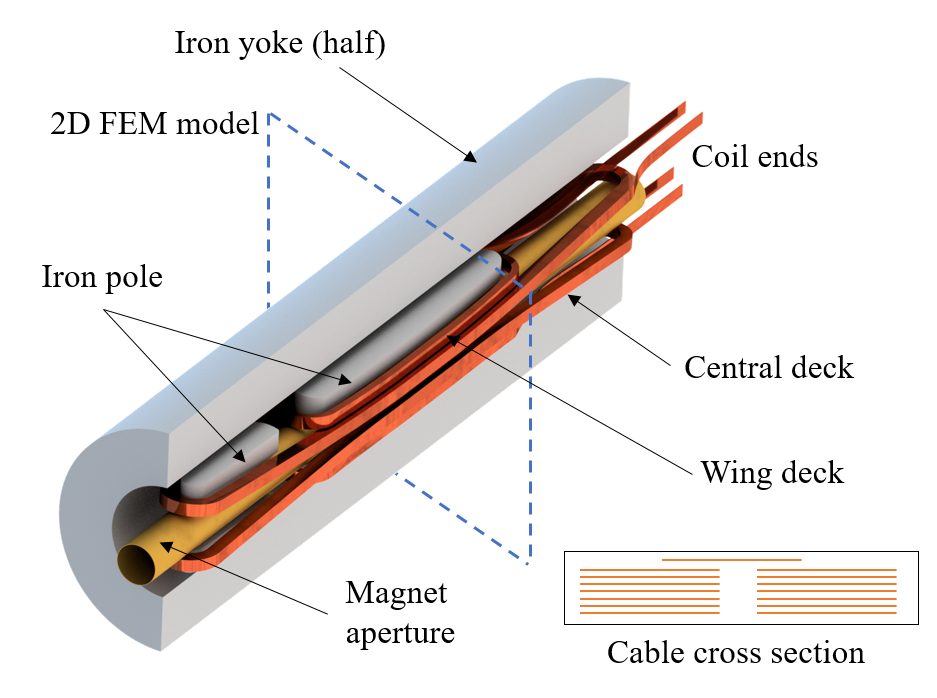}
	\caption{Simplified rendering of the \FEATHER{} magnet. The coil is composed by two pairs of central and wing decks. The cable is made of 15 tapes fully transposed with the Roebel technique. The cross-section of the cable is shown in the lower-right corner. The magnetic circuit is composed by four iron poles and a cylindrical iron yoke (half-shown).}
	\label{FIG_AssemblyFeatherM2}
\end{figure}
\begin{figure}[tb]
  \centering
	\includegraphics[width=8.0cm]{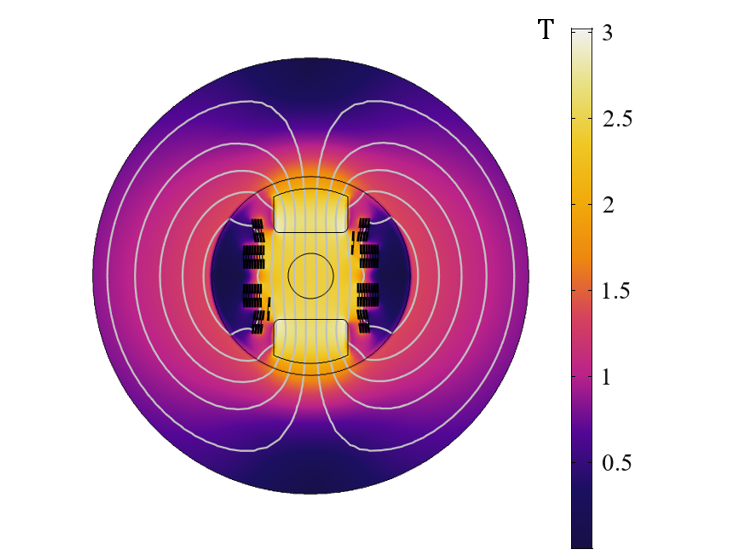}
	\caption{Magnetic field in T, at $\SI{5}{\kilo\ampere}$ and $\SI{4.5}{\kelvin}$, shown for the 2D cross-section of the \FEATHER{} magnet. The peak magnetic field reached in the aperture is about $\SI{2.5}{\tesla}$. }
	\label{FIG_FeatherM2MagneticField}
\end{figure}
The FEM model of the \FEATHER{} dipole magnet refers to the magnet version {{M}.1-2}, which is wound using a coated conductor produced by Sunam~\cite{SUNAM2019website}. The geometric and superconducting properties of the tape are reported in Table~\ref{TAB_TapeSpecifications}. This particular tape limits the magnet current to $\SI{5}{\kilo\ampere}$ and the peak field in the aperture to $\SI{3}{\tesla}$. A simplified rendering of the magnet is given in Fig.~\ref{FIG_AssemblyFeatherM2}, where for the sake of clarity, only the components relevant for the numerical analysis are shown. The coil is composed by two poles, each made of two windings named central and wing decks, and is designed to optimize the tape-field alignment~\cite{van2016measurement}. The magnetic field is shaped in the magnet aperture by means of iron poles. The outer iron yoke intercepts the stray field and allows for operating the magnet in a stand-alone configuration. The central cross-section of the magnet is used as geometry input for the 2D FEM model. The magnetic field solution, in Tesla, is shown in Fig.~\ref{FIG_FeatherM2MagneticField} for a current of $\SI{5}{\kilo\ampere}$ at $\SI{4.5}{\kelvin}$. The key features and the relevant simplifications of the model are discussed in the remainder of this section.
\begin{table}[tb]
    \caption{\FEATHER{} tape specifications}
    \label{TAB_TapeSpecifications}
    \centering
    \TabTapeSpecifications
\end{table}

\subsection{Coil Geometry}
    \label{SUBSEC_CoilGeometry}
The model is implemented for a 2D transverse field configuration, thus neglecting the magnetic effects of the end-coils. Due to the presence of the layer jumps connecting the lower and the upper windings in the coil, the magnetic symmetry in the cross-section of the magnet is not preserved. For this reason, the model accounts for a four-quadrants geometry, including the layer jumps in the first and third quadrant. The layer jump is visible in Fig.~\ref{FIG_FeatherM2MagneticField}, as a cable slightly misaligned with respect to the coil decks. 

HTS tapes feature a multi-material and multi-layer structure. At the same time, the tape used in the coil has a width-to-thickness ratio of about two orders of magnitude. This justifies approximating the geometry of the tape with a line~\cite{krahenbuhl1993thin}. In this way, the discretization of the thickness of the superconductor is avoided. At the same time, the physical properties of the materials composing the tape are homogenized. Such simplification is adopted to ensure an acceptable computational time, as the 2D model accounts for 648 tapes over four quadrants.

\subsection{Current Sharing Approximation}
    \label{SUBSEC_CurrentSharingRegime}
The cable used in the coil is made of 15 tapes, which are fully transposed using the Roebel technique~\cite{goldacker2007roebel,goldacker2014roebel}. The cross-section of the cable used in the numerical  model is sketched in the box of Fig.~\ref{FIG_AssemblyFeatherM2}, where each line represents a tape. Each tape is electrically connected in a parallel configuration, allowing for the redistribution of the supply current. Moreover, the Roebel transposition enforces the same electrical impedance for each of the tapes composing the cable, providing an even current distribution. For this reason, the same fraction of the supply current is imposed in the numerical model to each of the tapes, excluding current redistribution phenomena. Coupling currents~\cite{wilson1983superconducting} are also excluded, since they represent a second-order effect with respect to persistent currents~\cite{van2016high}.

Within each tape, current sharing phenomena are modeled by means of an equivalent surface resistivity, which homogenizes the superconducting and normal-conducting layers, as detailed in~\cite{bortot2020coupled}. The surface resistivity depends from the power law in \eqref{EQ_powerLaw}, thus is affected by the ${n}$-value. From magnet measurements~\cite{van2018powering}, an ${n}$-value of 5 was experimentally found, outside the expected range of 20-30~\cite{ghosh2004v}, and attributed to unbalanced tape joints.  However, note that for persistent magnetization the local critical current density is the relevant quantity and so the joint resistance is not the relevant quantify for calculating the persistent magnetization. Unfortunately, the tape was not characterized individually and so the uncertainty of the superconducting properties of the tapes is significant and the ${n}$-value is not known. To overcome this issue, a parametric sweep is performed for $4\leq{n}\leq30$, quantifying the sensitivity of the model. The results are compared with measurements in Section~\ref{SEC_SimWithMeas}.

\subsection{Critical Current Density Fit}
    \label{SUBSEC_CriticalCurrentDensityFit}
%
\begin{table}[tb]
    \caption{Parameters used for the $\SCAl{J}{\RM{c}}$ fit}
    \label{TAB_IcritFitParameters}
    \centering
    \TabIcFitParameters
\end{table}
\begin{figure}[tb]
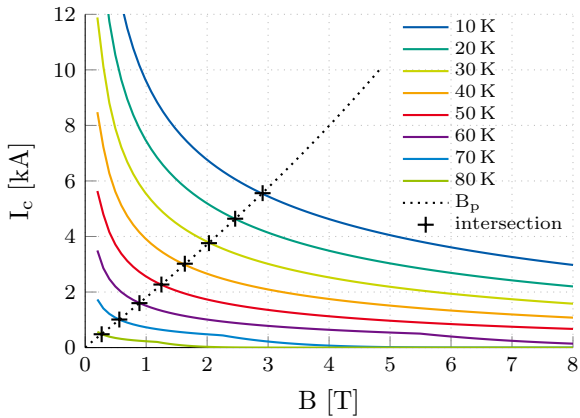

    \centering
    \FigIcFitFunB       
    \caption{Calculation of the critical current in the \FEATHER{} magnet as a function of the magnetic field. The critical current is obtained for each temperature as the intersection point (markers) of the magnetic characteristic of the magnet (dotted line), known also as the load line, with the critical current provided by the fit (solid lines), assuming a perpendicular magnetic field to the cable.}
    \label{FIG_IcFitFunB}
\end{figure}
\begin{figure}[tb]
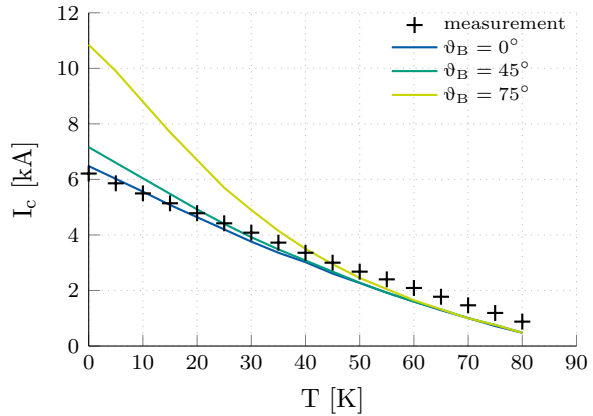

    \centering
    \FigIcFitFunT       
    \caption{Calculated critical current of the cable as a function of temperature, parametrized by the magnetic field angle with respect to the cable perpendicular direction. The markers show the measured critical current in the \FEATHER{} magnet.}
    \label{FIG_IcFitFunT}
\end{figure}
\begin{figure}[tb]
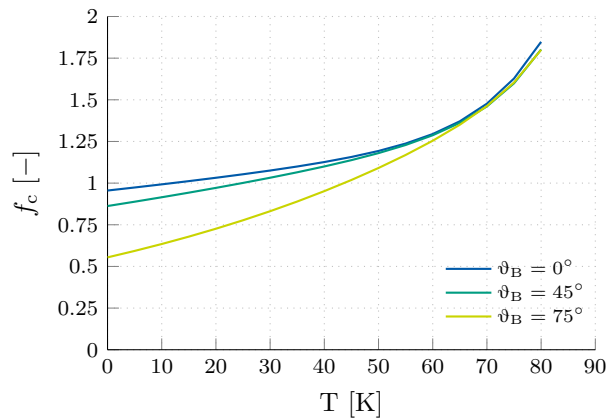

    \centering
    \FigIcFitLiftFc       
    \caption{Correction factor applied to the critical current fit, as a function of temperature, parametrized by the magnetic field angle with respect to the cable perpendicular direction.}
    \label{FIG_IcFitLiftFc}
\end{figure}
The critical current density $\SCAl{J}{\RM{c}}$ in~\eqref{EQ_powerLaw} affects the persistent currents dynamics and, ultimately, the field quality in the magnet. In {R}eBCO tapes, $\SCAl{J}{\RM{c}}$ shows an anisotropic, field- and temperature-dependent behavior, as $\SCAl{J}{\RM{c}}(\SCA{B},\SCA{T},\SCAl{\RMtheta}{\RM{B}})$, where $\SCAl{\RMtheta}{\RM{B}}$ is the magnetic field angle with respect to the direction perpendicular to the tape wide surface. 

The behavior of $\SCAl{J}{\RM{c}}$ is included in the model by means of the numerical fit provided in~\cite{fleiter2014characterization}. The fit parameters, reported in Table~\ref{TAB_IcritFitParameters}, are taken from~\cite{van2016high}, since no data was available for the used Sunam tape. For this reason, the fit is scaled in order to provide a critical current for the \FEATHER{} coil which is consistent with measurements~\cite{van2018powering}, as follows.

The magnetic characteristic of the magnet, known also as the load line, is calculated numerically by means of magnetostatic simulations. With respect to Fig.~\ref{FIG_IcFitFunB}, the load line is given in terms of peak magnetic field $\SCAl{B}{\RM{p,coil}}$ in the coil as a function of the supply current (dotted line). The critical current is then given for each temperature as the intersection of the load line (markers) with the critical current provided by the fit (solid lines). The magnetic field is assumed to be perpendicular to the cable, as $\SCAl{\RMtheta}{\RM{B}}=0^{\circ}$. In Fig.~\ref{FIG_IcFitFunT}, the calculated critical current is compared with the measurements, and parametrized by the field angle. The assumption of field perpendicularity gives the best agreement with the measured data. The fitting factor is finally obtained as
    \begin{equation}
        \label{EQ_defCorrFactorJc}
            \EQdefCorrFactorJc{},        
    \end{equation}
where $\SCAl{I}{\RM{c,meas}}$ is the critical current obtained from measurements, and $\SCAl{S}{\RM{HTS}}$ is the superconducting cross-section of the cable. The fitting factor is shown as a function of temperature in Fig.~\ref{FIG_IcFitLiftFc}, and parametrized by the field angle. The factor obtained for $\SCAl{\RMtheta}{\RM{B}}=0^{\circ}$ is used in the model for scaling the critical current density fit.

\subsection{Iron Hysteresis}
    \label{SUBSEC_IronHysteresis}
%
\begin{figure}[tb]
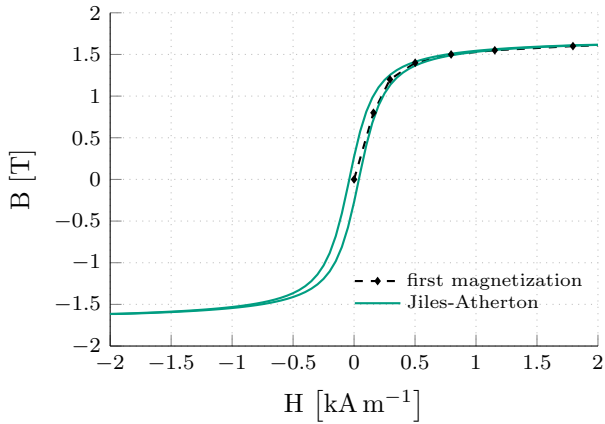

    \centering
    \FigIronHysteresis      
    \caption{Nonlinear magnetic characteristics of the iron used in the \FEATHER{} model, represented with: 1) the first magnetization curve, and 2) the hysteresis loop provided by the Jiles-Atherton model.}
    \label{FIG_IronHysteresis}
\end{figure}
\begin{table}[tb]
    \caption{Parameters for the Jiles-Atherton hysteresis model}
    \label{TAB_JilesAthertonParameters}
    \centering
    \TabJilesAthertonParameters
\end{table}
\begin{figure}[tb]
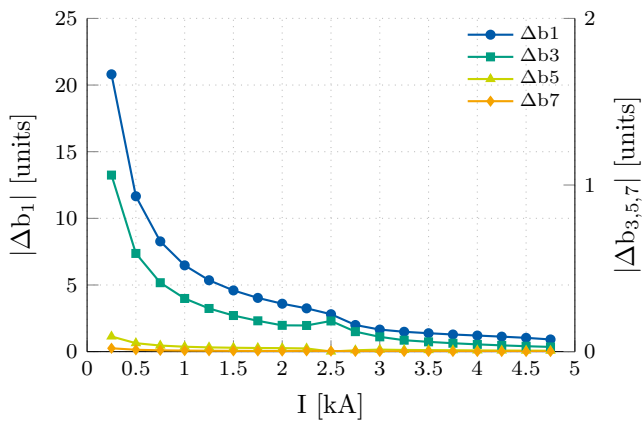

    \centering
    \FigIronHysteresisMultipoles 
    \caption{Amplitude of the magnetization loop of the field multipoles, as a function of current. The magnitude of $\RMDelta\SCA{b1}$ is given by the left axis, whereas the magnitude of $\RMDelta\SCA{b3}$, $\RMDelta\SCA{b5}$ and $\RMDelta\SCA{b7}$ is given by the right axis.}
    \label{FIG_IronHysteresisMultipoles}
\end{figure}
The magnetic material used in the yoke of the Feather-M2 magnet is chosen to minimize the detrimental influence of the iron hysteresis on the magnetic field quality. No material characterization data was available, however the iron is similar to the one of the LHC main dipoles. For this reason, the numerical model considers the same nonlinear $\RM{B(H)}$ curve used for the {LHC}~\cite{russenschuck1993roxie}. The curve is shown as a dashed line in Fig.~\ref{FIG_IronHysteresis}. In stand-alone operations, most of the outer iron yoke of the \FEATHER{} magnet remains unsaturated up to the maximum current of $\SI{5}{\kilo\ampere}$. For this reason, the contribution of the iron hysteresis on the field quality cannot be neglected a priori, and is estimated as follows.

A coercive field $\SCAl{H}{\RM{c}}$ of $\SI{40}{\ampere\per\meter}$ is assumed for the material used in the yoke, in accordance with the {LHC} specifications ($\SCAl{H}{\RM{c}}\leq\SI{60}{\ampere\per\meter}$~\cite{russenschuck2011field}). The iron hysteresis is included by using the Jiles-Atherton model~\cite{jiles1986theory}, whose loop is determined by using the B(H) curve as reference, and shown in Fig.~\ref{FIG_IronHysteresis}. The relevant parameters for the hysteresis model are obtained using the open-source algorithm from~\cite{szewczyk2018open} and are reported in Table~\ref{TAB_JilesAthertonParameters}. 

The hysteresis contribution is calculated on a simplified \FEATHER{} model, including only the iron as nonlinear effect. The multipole coefficients are obtained as function of the current for both the upper and the lower hysteresis curve, then the two data sets are compared. Their difference $\RMDelta\SCA{b}$ provides the amplitude of the magnetization loop for each coefficient, giving an estimation for the effect of the iron hysteresis on the field quality. 

With respect to Fig.~\ref{FIG_IronHysteresisMultipoles}, the hysteresis of the iron has a minor influence at low current on the main field component $\SCAl{b}{1}$. A peak value of about 20 units is found and it rapidly decreases once the current is increased, since the with of the hysteresis loop narrows. Concerning the higher order multipoles $\SCAl{b}{3}$, $\SCAl{b}{5}$ and $\SCAl{b}{7}$, there is almost no influence since the contribution is always less than one unit. As a consequence, the contribution to the field from the interaction between the iron hysteresis and the screening currents in the coil can be reasonably assumed a second order effect, thus negligible.

The analysis shows a limited influence from the iron hysteresis on the magnetic field quality, at the price of an increased computational cost. For this reason, the iron hysteresis is excluded from the numerical model of the \FEATHER{} magnet.

\section{Comparison of Simulations with Measurements} 
    \label{SEC_SimWithMeas}
The numerical model of the \FEATHER{} magnet is validated by comparing the simulation results of the magnetic field quality in the magnet aperture with available experimental observations. The comparison is done for four scenarios, which differ in the peak current $\SCAl{I}{\RM{p}}$ (i.e., peak magnetic field) and operational temperature $\SCAl{T}{\RM{op}}$ of the magnet. The relevant parameters characterizing the scenarios are reported in Table~\ref{TAB_Scenarios}. It is worth noting that as $\SCAl{T}{\RM{op}}$ is increased, $\SCAl{I}{\RM{p}}$ is reduced accordingly, such that the ratio between the peak current and the critical current of the cable is kept constant. In accordance with measurements, $\SCAl{T}{\RM{op}}$ is assumed as homogeneous and constant in the numerical model, for each scenario. 

In the following, the measurement and simulation setups are discussed, and the comparison between experimental and numerical results is presented. All the simulations are carried out on a standard workstation (Intel$^{\circledR}$ Core i7-3770 {CPU} $@$ $\SI{3.40}{\giga\hertz}$, $\SI{32}{\giga\byte}$ of {RAM}, Windows-$10^{\circledR}$ Enterprise 64-bit operating system), using the proprietary {FEM} solver {COMSOL} {M}ultiphysics$^{\circledR}$~\cite{comsol2005comsol}.

\begin{table}[tb]
    \caption{Simulated scenarios: main parameters}
    \label{TAB_Scenarios}
    \centering
    \TabAnalysisScenarios
\end{table}

\subsection{Measurement Setup}
    \label{SEC_MeasurementSetup}
%
\begin{figure}[tb]
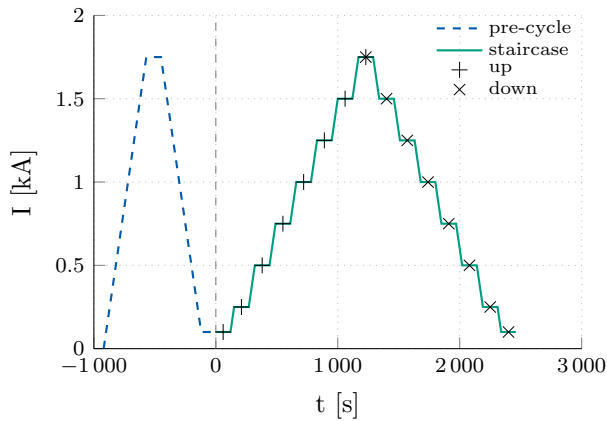

  \centering
    \FigStaircase
	\caption{Example of a current profile used in the simulations (scenario at $\SI{1.75}{\kilo\ampere}$ and $\SI{68}{\kelvin}$). The current follows a trapezoidal pre-cycle, then a staircase profile, up to the peak current and back. The markers at the current plateaus ($up$ and $down$ labels) represent the evaluation points for the magnetic field quality for both the ascending and descending part of the staircase.}
	\label{FIG_NormalizedStaircase}
    \vskip -0.25cm
\end{figure}   
Rotating-coil magnetometers, also known as harmonic coils, are electromagnetic transducers for measuring the $\SCAl{B}{k}$ and $\SCAl{A}{k}$ field multipoles. The coil shaft is positioned parallel to the magnetic axis of the magnet, and it is rotated in the magnet aperture. The change of flux linkage $\RMPhi$ induces, by integral Faraday\'{}s law $\SCAl{U}{\RM{m}} = -\RM{d}_{t}{\RMPhi}$, a voltage signal $\SCAl{U}{\RM{m}}$ which is measured at the terminals of the coil. By integrating in time the voltage signal, the flux linkage is obtained and given as a function of the series expansion of the radial field~\cite{russenschuck2011field}. Assuming a coil of negligible thickness, perfectly centered in the aperture of a magnet, and rotating with angular velocity $\RMomega$, then for an arbitrary angle $\ITvarphi(t)=\RMomega t + \ITvarphi_{0}$ the flux linkage is given at time $t$ as
\begin{align}
    \label{EQ_defRotCoilPhi} 
        \EQdefRotCoilPhi{}, \\
    \label{EQ_defRotCoilSensFactor} 
        \EQdefRotCoilSensFactor{},    
\end{align}
where the coil sensitivity factor $\SCAl{f}{\RM{s}}(k)$ embeds the coil geometric parameters, namely the number of turns $\SCAl{N}{\RM{c}}$, longitudinal length $\SCAl{l}{\RM{c}}$ and the mean radius $\SCAl{r}{\RM{c0}}$. Such parameters are calibrated in a dipole and quadrupole reference magnet.

Encouraged by the results obtained from the flux sensors presented in~\cite{petrone2018measurement}, a dedicated rotating-coil magnetometer was developed and employed to test the \FEATHER{} magnet in the variable temperature cryostat at CERN. The constructed coil shaft is composed of a chain of five Printed-Circuit Boards (PCBs), ($\SI{200}{mm}$ in length and~$\SI{35}{mm}$ in width), that span the entire magnet length including the fringe-field areas. Every PCB board contains three coils mounted radially, with an active surface of~$\SI{0.1817}{m^{2}}$. 

For the magnetic-field harmonics, the measurement sensitivity is improved by connecting two coils in anti series; for the dipole magnet measurement the external coil minus the central coil. {CERN} proprietary digital cards~\cite{arpaia2012performance} integrate the induced voltages in the coils rotating at a frequency of~$\SI{2}{Hz}$. In this paper, the measurement results are taken from the longitudinal center of the magnet (the central element of the rotating shaft of 200 mm in length), delivering a measurement precision of a magnetic-field harmonic of $\pm0.05$ units.

\subsection{Simulation Setup}
    \label{SEC_SimulationSetup}
To match the experimental procedure, a current excitation is applied as a source for the numerical model. With respect to the example provided in Fig.~\ref{FIG_NormalizedStaircase}, the current follows firstly a trapezoidal pre-cycle, then a staircase profile spanning from a minimum value of $\SI{0.25}{\kilo\ampere}$ up to the peak current, and back. The aim of the pre-cyle is to remove the dependency of the superconducting coil on the first magnetization cycle. The staircase signal is composed of steps of steepness $\SI{10}{\ampere\per\second}$, which increase the current by $\RMDelta\SCA{I}=\SI{250}{\ampere}$, and then keep it constant for $\RMDelta\SCAl{t}{\RM{flat}}=\SI{120}{\second}$. For each midpoint in the staircase plateaus, sowed in Fig.~\ref{FIG_NormalizedStaircase} with a marker, the magnetic field quality is calculated and compared with measurements. The number of steps is adapted for each scenario, in order to reach the prescribed peak current. The shape of the current excitation and the evaluation points for the field quality are consistent with the ones used in the measurements.

Following~\cite{petrone2018measurement}, the staircase current profile is used to quantify the influence of hysteresis phenomena occurring within both the superconducting coil and the iron yoke of the magnet, as follows. With respect to Fig.~\ref{FIG_NormalizedStaircase}, for each current step the field quality is measured and simulated twice, once during the ramp-up and then during the ramp-down, and the results are grouped in pairs. Subsequently, the difference of the field multipoles is calculated for each pair of field quality evaluations. Thanks to this operation, the contributions of the non-ideal geometry of the coil and the iron saturation are canceled out, being the same for both evaluations in each pair, and the residual is attributed to the hysteresis phenomena. Since the iron hysteresis was previously found to produce only a second-order effect on the field quality (see Section~\ref{SUBSEC_IronHysteresis}), the hysteresis contribution is fully attributed to the persistent magnetization of the superconducting coil.


\subsection{Results}
    \label{SEC_Results}
%
\FigureResultPlots{}
The measured and simulated field multipole coefficients are given in Fig.~\ref{FIG_FieldQuality}. The markers represent the measurements which are split in the $up$ and $down$ datasets, accordingly to the upward and downward part of the current staircase (see Fig.~\ref{FIG_NormalizedStaircase}). The shaded area gives the envelope of the numerical solutions obtained by a parametric sweep of the ${n}$-value between 4 and 30. As an example, the simulation results for ${n}=20$ are highlighted with a solid line. The dashed line represents the ideal case in which the the screening currents do not have any influence on the field quality. This is obtained by artificially increasing the resistivity of the superconducting coil until a homogeneous current density distribution is achieved in the time domain simulation. The rows show, from top to bottom, the normal dipole field $\SCAl{B}{\RM{1}}$ and the multipoles $\SCAl{b}{\RM{3}}$, $\SCAl{b}{\RM{5}}$ and $\SCAl{b}{\RM{7}}$, as a function of the source current. 
The columns separate the results by the operational temperature of the magnet, namely 4.5, 9, 25, and $\SI{68}{\kelvin}$ or, in other words, the simulated scenario. 

The field multipoles keep qualitatively the same behavior through the different scenarios (see Fig.~\ref{FIG_FieldQuality}, row by row). Moreover, the $\SCAl{b}{\RM{3}}$ and $\SCAl{b}{\RM{5}}$ multipoles are reduced as the the current is increased. The $\SCAl{b}{\RM{7}}$ coefficient is negligible with respect to the others. The scenario at $\SI{4.5}{\kelvin}$ shows the highest variation in the magnitude of the multipoles. At low current, the $\SCAl{b}{\RM{3}}$ contribution increases of about a factor 2, from 200 to 400 units, and the $\SCAl{b}{\RM{5}}$ multipole shows an increase of a factor 8, from 10 to 80 units. This might be explained as screening currents are higher at low temperature, due to the higher critical current density of the tape. 

Hysteresis phenomena in the \FEATHER{} magnet create the magnetization loops which are present in the measured and simulated data sets. The loops are found to be at least one order of magnitude smaller than the absolute value of the multipole coefficients. For this reason, the width of the loops is shown separately in Fig.~\ref{FIG_FieldQualityPersistentCurrent}. The layout and the meaning of symbols is the same as before for Fig.~\ref{FIG_FieldQuality}. The rows show from top to bottom the variation in units for the multipoles $\SCAl{b}{\RM{1}}$, $\SCAl{b}{\RM{3}}$, $\SCAl{b}{\RM{5}}$ and $\SCAl{b}{\RM{7}}$, as a function of the supply current. The columns separate the results by the operational temperature of the magnet, namely 4.5, 9, 25, and $\SI{68}{\kelvin}$. 

The width of the magnetization loops due to persistent currents does not exceed twenty units for $\SCAl{b}{\RM{1}}$ and $\SCAl{b}{\RM{3}}$, two units for $\SCAl{b}{\RM{5}}$ and one unit for $\SCAl{b}{\RM{7}}$. The trend is generally monotone, showing the multipoles decreasing as the current increases, and vanishing as the current reaches its peak value. The $\SCAl{b}{\RM{1}}$ coefficient is an exception, as it has a peak around $\SI{3.5}{\kilo\ampere}$, when the pole of the iron yoke saturates. In the ideal case where the screening currents are neglected, the width of the magnetization loops is always zero.

\section{Discussion} 
    \label{SEC_Discussion}
The field quality in the \FEATHER{} magnet shows $\SCAl{b}{\RM{3}}$ and $\SCAl{b}{\RM{5}}$ coefficients which are much higher than the few units typically required by accelerator quality standards~\cite{russenschuck2011field} (see Fig.~\ref{FIG_FieldQuality}). This might be explained by the influence of the outer iron yoke which is not yet optimized for field quality purposes. The field error is governed by the $\SCAl{b}{\RM{3}}$ coefficient, whereas $\RM{b_{5}}$ is about one order of magnitude smaller, and $\RM{b_{7}}$ is negligible. 

The magnet design is optimized to deliver the highest field quality when operating in nominal conditions. As a consequence, for an increasing supply current (i.e., increasing main dipole field), the multipole coefficients are decreasing. If the temperature is increased, the peak supply current needs to be reduced accordingly, to cope with the temperature dependency of the cable critical current. The working point of the magnet then shifts from nominal conditions, and the $\SCAl{b}{\RM{3}}$ and $\SCAl{b}{\RM{5}}$ multipole coefficients increase.

Referring to Fig.~\ref{FIG_FieldQualityPersistentCurrent}, the contribution of the screening current-induced magnetic field to the field quality never exceeds 20 units, thus it is one order of magnitude smaller than the total field error (see Fig.~\ref{FIG_FieldQuality}). The numerical analysis gave better agreement with measurements for high ${n}$-values ($\geq$20), whereas for small ${n}$-values ($\leq$10), the contribution of the persistent currents is overestimated. The results seem to confirm that the low quality of the tape (measured ${n}$-value of 5) is due to the tape joints, which do not play any role in the dynamics of the persistent currents. The limited contribution of the persistent magnetization might be explained with the coil design, which is optimized to align the tapes with the magnetic field lines~\cite{van2016measurement}, limiting the flux linked to the surface of the tapes, and thus magnetization phenomena.

\begin{figure}[tb]
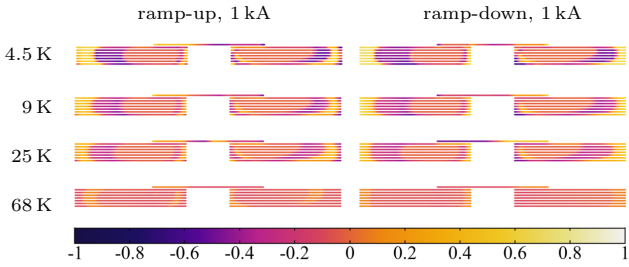

  \centering
    \FigJnormDistributionInCable
	\caption{Current density distribution in the most inner turn of the upper deck, normalized with the critical current density at zero field and $\SI{4.5}{\kelvin}$ $\SCAl{J}{\RM{crit,0}}=\SI{138}{\kilo\ampere\per\milli\metre\squared}$. The distribution is given at $\SI{1}{\kilo\ampere}$ for both, the ramp-up and the ramp-down, for different temperatures.}
	\label{FIG_JnormDistributionInCable}
\end{figure}

By increasing the operational temperature of the magnet, the critical current density of the tape is reduced, leading to a faster field diffusion in the tape, and consequently to a more homogeneous current density distribution in the cable. This is shown in Fig.~\ref{FIG_JnormDistributionInCable} where the current density distribution normalized to $\SCAl{J}{\RM{c}}(\SI{4.5}{\kelvin},\SI{0}{\tesla},\SI{0}{\degree})=\SI{138}{\kilo\ampere\per\milli\metre\squared}$ is given for the most inner turn of the upper deck. As the supply current is increased, the persistent currents tend to vanish independently from the operational temperature. This might be explained by the saturation of the tape due to the supply current. 

Numerical simulations are in agreement with measurements, consistently reproducing both the magnetic overall field quality and the persistent magnetization contribution. Still, simulation results are affected by the uncertainty on the superconducting properties of the tapes used in the \FEATHER{} magnet. Nevertheless, the analysis is relevant as it clearly shows which properties are important for understanding the field-quality-behavior of HTS accelerator magnets. For this reason, a more extensive tape characterization is recommended for future magnets, thus reducing the uncertainty in the material properties and enhancing the confidence and accuracy in dynamic field quality simulations.

\section{Conclusions and Outlook} 
    \label{SEC_Conclusions and Outlook}
This paper presents the time-domain analysis of the demonstrator magnet \FEATHER{}, an HTS insert dipole designed to provide an additional $\SI{5}{\tesla}$ in the $\mathrm{Nb_{3}Sn}$ {FRESCA}2 background magnet, up to peak fields of $\SI{18}{\tesla}$ in the magnet aperture. The analysis quantifies the influence of the screening current-induced magnetic field on the magnetic field quality in the magnet aperture. Simulations reproduce the powering cycle of the magnet for different temperatures and operating currents by using a staircase-shaped current profile. The magnet is simulated in a stand-alone configuration, such that numerical results are verified with available measurements.

For this case study, the field quality error due to persistent magnetization phenomena affects mostly the main field component, and it is limited to 20 units. Moreover, the error is significantly reduced once the supply current is increased to the operational value, saturating the tape. The coupling of the scrrening currents with the hysteresis of the iron is found to be negligible. Thus, the aligned-coil design might be a key-feature for ensuring accelerator quality standards in the magnetic field of future HTS accelerator magnets.

The numerical analysis is carried out under magnetoquasistatic assumptions, using time-domain simulations based on a coupled $\VEC{A}$-$\VEC{H}$ formulation implemented in a 2D FEM model. The formulation is verified against analytical solutions from previous literature, and the model is validated with available experimental data. The model requires only one scalar correction parameter for the power law, compensating for the uncertainty in the critical current density of the tape. Simulations quantify the influence of the coil electrodynamics on the magnetic field, achieving satisfactory agreement with measurements. The computational time is less than one hour for each simulation, on a standard workstation. The accuracy of the model may be increased by a better knowledge of both the critical surface current of the tape used for the coil, and the magnetization curve of the iron used for the yoke.

The model provides for each tape an accurate quantification of the dynamic distribution of the persistent currents, which can be used not only for the magnetic field quality analysis, but also for the calculation of the Joule losses and the dynamic forces in the coil. As screening currents provide the principal contribution to dynamic losses in HTS tapes, such valuable insights can be integrated for the future design of HTS magnets, e.g. within a numerical optimization workflow for quench protection studies.

%% file: Section/SEC03_Acknowledgements.tex
\section{Acknowledgments} 
    \label{Acknowledgements}
    
This work has been sponsored by the Wolfgang Gentner Programme of the German Federal Ministry of Education and Research (grant no. 05E15CHA), by the ‘Excellence Initiative’ of the German Federal and State Governments and by the Graduate School of Computational Engineering at Technische Universit{\"a}t Darmstadt. Parts of the work have been funded by the BMBF project “Diagnose of high-intensity hadron beams (DIAGNOSE)”, under grant no. 05P18RDRB1.

The authors would like to acknowledge the fruitful collaboration between CERN and the Technische Universit{\"a}t Darmstadt, within the framework of the STEAM collaboration project~\cite{STEAM2019website}. The authors would like to thank S. Russenschuck for the constructive comments on the paper. The authors would also like to thank J. Murtomaki, T. Nes and S. Richter for fruitful discussions and valuable suggestions concerning the dynamics of HTS tapes.